\documentclass{article}
\usepackage{fontspec}%
\usepackage{iftex}
\ifluatex
\newfontfamily\arabicfont[
  Path=fonts/,
  Script=Arabic,
  Scale=1.0,
  Renderer=HarfBuzz,
  RawFeature={+ss05,+ss06,+ss07}
]{Amiri-Regular.ttf}
\else
\newfontfamily\arabicfont[
  Path=fonts/,
  Script=Arabic,
  Scale=1.0,
  RawFeature={+ss05,+ss06,+ss07}
]{Amiri-Regular.ttf}
\fi

\ifluatex
\protected\def\arb#1{{\arabicfont\textdir TRT\pardir TRT #1}}
\else
\protected\def\arb#1{\RLE{\arabicfont#1}}
\fi

\usepackage{hyperref}

\usepackage[preprint]{icml2026}

\usepackage{amsmath,amssymb,amsfonts}%
\usepackage{graphicx}%
\usepackage{textcomp}%
\usepackage{array}%
\usepackage{booktabs}%
\usepackage{tabularx}%
\usepackage{multirow}%
\usepackage{longtable}%
\usepackage{tikz}%
\ifxetex
\makeatletter
\providecommand\UseMathForPositioningText{}%
\providecommand\@kernel@tabular@init{}%
\makeatother
\usepackage{bidi}
\fi

\icmltitlerunning{Automatic Pronunciation Error Detection and Correction of the Holy Quran's Learners Using Deep Learning}

\begin{document}

\twocolumn[
  \icmltitle{Automatic Pronunciation Error Detection and Correction of the Holy Quran's Learners Using Deep Learning}

  \icmlsetsymbol{equal}{*}

  \begin{icmlauthorlist}
    \icmlauthor{Abdullah Abdelfttah}{equal,asu}
    \icmlauthor{Mahmoud I. Khalil}{asu}
    \icmlauthor{Hazem Abbas}{asu}
  \end{icmlauthorlist}

  \icmlaffiliation{asu}{Faculty of Engineering, Ain Shams University, Cairo, Egypt}

  \icmlcorrespondingauthor{Abdullah Abdelfttah}{quran.muaalem@gmail.com}
  \icmlcorrespondingauthor{Mahmoud I. Khalil}{mahmoud.khalil@eng.asu.edu.eg}
  \icmlcorrespondingauthor{Hazem Abbas}{hazem.abbas@eng.asu.edu.eg}

  \icmlkeywords{Automatic Pronunciation Error Detection, Tajweed, Holy Quran, Deep Learning, CTC}

  \vskip 0.3in
]

\printAffiliationsAndNotice{}

\fancyfoot[L]{\footnotesize \textit{Preprint}}
\fancyfoot[C]{\thepage}

\begin{abstract}
Assessing spoken language is challenging, and quantifying pronunciation metrics for machine learning models is even harder. However, for the Holy Quran, this task is enabled by the rigorous recitation rules (Tajweed) established through the efforts of Muslim scholars, making highly effective assessment possible. Despite this advantage, the scarcity of high-quality annotated data remains a significant barrier. In this work, we bridge these gaps by introducing: (1) A 98\% automated pipeline to produce high-quality Quranic datasets -- encompassing collection of recitations from expert reciters, segmentation at pause points (waqf) using our fine-tuned wav2vec2-BERT model, transcription of segments, and transcript verification via our novel Tasmeea algorithm; (2) 848 hours of audio (286K annotated utterances); (3) qdat\_bench, a benchmark covering phonemes, diacritization, and Tajweed rules (Ghunnah, Qalqalah, Madd) on real recitation errors containing 159 samples; (4) A novel ASR-based approach for pronunciation error detection utilizing our custom Quran Phonetic Script (QPS) to encode Tajweed rules (unlike the IPA standard for Modern Standard Arabic). QPS uses an 11-level script: phoneme level (encoding Arabic letters with short/long vowels) and sifat level (encoding articulation characteristics of every phoneme). We further present comprehensive modeling with our novel multi-level CTC model, which achieved 0.21\% and 1.94\% average Phoneme Error Rate (PER) on the test set and qdat\_bench respectively, with a 75.8\% Tajweed F1 score. We release our work as open-source: \href{https://obadx.github.io/quran-muaalem/en/}{https://obadx.github.io/quran-muaalem/en/}

\end{abstract}

\section{Introduction}
\noindent
Assessing pronunciation is not a simple task \cite{kheir2023automatic}, as it relies not only on correct phoneme pronunciation but also on intonation, prosody, and stress. The Holy Quran presents unique characteristics: it is among the easiest spoken texts to learn despite containing special phonemes absent in other languages.

The Holy Quran is considered the word of Allah by Muslims. It was revealed to Prophet Muhammad (peace be upon him) and holds immense significance for the Muslim community. The Quran possesses unique linguistic characteristics; it is neither poetry nor prose in the traditional sense of Arabic literature but introduces a new art form that takes on a lyrical quality through recitation (Tajweed). Consequently, Quranic recitation involves unique pronunciation characteristics, such as elongation (Madd), nasalization (Ghunnah), and echo (Qalqalah). According to Ibn al-Jazari, these patterns must be pronounced consistently~\cite{AlJazariyyahSwaid}. As non-Arabic speakers convert to Islam, pronunciation errors can occur. To preserve the exact pronunciation and meaning, ancient Muslim scholars established recitation rules derived from the Quran recitation itself.

The pronunciation of the Holy Quran is governed by rigorously strict rules formally defined by ancient Muslim scholars since the 6th century. Despite being precise and accurate, these rules have not been comprehensively digitized (to our knowledge) for Quranic pronunciation assessment.

This paper focuses on automatic detection and localization of Tajweed Rules and Sifat-related pronunciation errors from audio, enabling formative feedback, automated tutoring, and standardized assessment.

To the best of our knowledge, RDI \cite{sherif2007enhancing} was the first to apply AI for detecting pronunciation errors alongside Tajweed rules. However, they did not disclose their phoneticization process nor release their data or models publicly. Subsequent research has remained largely incomplete -- often focusing on a subset of rules without open-source code or datasets. To address these gaps, we introduce:
\begin{itemize}
    \item A phonetizer that encodes all Tajweed rules and articulation attributes (Sifat) defined by classical scholars (except Ishmam).
    \item A 98\% automated pipeline for generating highly accurate datasets from expert recitations.
    \item A dataset of ~286K annotated utterances (848 hours).
    \item qdat\_bench, a benchmark for phonemes, diacritization, and Tajweed rules (Ghunnah, Qalqalah, Madd) on real recitation errors (159 samples).
    \item A multi-level CTC model that demonstrates the learnability of QPS (0.21\% average PER).
\end{itemize}

Unlike previous approaches that use IPA-based phoneme sets for Modern Standard Arabic \cite{omran2023automatic}, diacritized Uthmani text without Tajweed rules \cite{khan2021tarteel}, or focus on single rules like Qalqalah only \cite{10092350}, our QPS comprehensively encodes all Tajweed rules and Sifat attributes in a unified representation.

Our multi-level CTC architecture offers significant token reduction compared to flat representations. A single-level approach encoding all Tajweed rules would require 99,072 tokens (cross-product of 43 phonemes × 10 Sifat × variations), whereas our hierarchical approach with 11 levels as shown in table~\ref{tab:vocabulary_size} achieves the same representational capacity with dramatically fewer output units, enabling more efficient training and better generalization.

\begin{table}[t]
	\centering
	\caption{Vocabulary size per level in the Quran Phonetic Script.}
	\label{tab:vocabulary_size}
	\begin{tabular}{|l|c|}
		\hline
		\textbf{Level} & \textbf{Vocabulary Size} \\
		\hline
		Phonemes & 43 \\
		Tafkheem / Tarqeeq & 3 \\
		Shidda / Rakhawa & 3 \\
		Moraqaq / Muftah & 2 \\
		Itbaq & 2 \\
		Safeer & 2 \\
		Qalqala & 2 \\
		Tikraar & 2 \\
		Tafashie & 2 \\
		Istitala & 2 \\
		Ghunnah & 2 \\
		\hline
	\end{tabular}
\end{table}

The paper is organized as follows: Related Work, Quran Phonetic Script, Data Pipeline, Modeling, Results, Limitations and Future Work, Conclusion, and Appendix.

\section{Related Work}

\subsection{Quran Pronunciation Datasets}
Table~\ref{tab:datasets-comparison} summarizes the most relevant Quran pronunciation datasets. EveryAyah is the largest openly available dataset with 26 complete \emph{Moshafs} segmented and annotated Ayah-by-Ayah by both experts such as Al Hossary and non-experts such as Fares Abbad. Qdat \cite{osman2021qdat} contains 1,509 utterances of single specific Ayahs labeled for three rules: Madd, Ghunna, and Ikhfaa. Although the scale is relatively small, it was widely adopted by the community \cite{10092350,shaiakhmetov2025evaluation} due to being open-source. The Tarteel v1 dataset \cite{khan2021tarteel} consists of 25K utterances with diacritics and no Tajweed rules. The latter is the Tarteel private dataset, a massive 9K-hour collection annotated with diacritics without Tajweed rules. The most recent benchmark is IqraaEval \cite{kheir2025towards}, which presents a test set of 2.2 hours from 18 speakers, but uses Modern Standard Arabic (MSA) without Tajweed rules.

None of the above provides explicit, rule-level Tajweed annotations across the full range of Hafs recitation rules. This scarcity has been a fundamental bottleneck.

\begin{table*}[t]
	\centering
	\scriptsize
	\begin{tabular}{p{2.8cm} p{2.3cm} p{2.2cm} p{3.5cm} c}
		\toprule
		\textbf{Dataset}                                                                                                                       & \textbf{Task}            & \textbf{Size}                           & \textbf{Tajweed Rules}                 & \textbf{Sifat} \\
		\midrule
		EveryAyah                                                                                                                              & ASR / Recitation         & 114 reciters                            & None                                   & No             \\
		Qdat \cite{osman2021qdat}                                                                                                              & Tajweed Detection        & 1,509 utterances                        & 3 (Madd, Ghunna, Ikhfaa)               & No             \\
		Tarteel v1 \cite{khan2021tarteel}                                                                                                      & Diacritization           & 25K utterances (68 h)                   & None                                   & No             \\
		Tarteel (private)                                                                                                                      & Diacritization           & 9K hours                                & None                                   & No             \\
		IqraaEval \cite{kheir2025towards}                                                                                                      & Pronunciation Assessment & 2.2 hours                               & None                                   & No             \\
		\textbf{muaalem-annotated-v3 (ours)}\footnote{\href{https://huggingface.co/datasets/obadx/muaalem-annotated-v3}{muaalem-annotated-v3}} & ASR + Tajweed Detection  & 848 h, 286K utterances, 22 reciters     & All Hafs Tajweed rules (except Ishmam) & Yes            \\
		\textbf{qdat\_bench (ours)}\footnote{\url{https://huggingface.co/datasets/obadx/qdat_bench}}                                                  & ASR + Tajweed Detection        & 159 samples (1 verse, Al-Ma'idah 5:109) & Madd, Ghunna, Ikhfaa, Qalqalah         & Yes            \\
		\bottomrule
	\end{tabular}
	\caption{Comparison of Quran pronunciation datasets. The Sifat column indicates whether the dataset explicitly encodes \emph{Sifat} (attributes of articulation of Holy Quran phonemes) at the phoneme level.}
	\label{tab:datasets-comparison}
\end{table*}

\subsection{Quran Pronunciation Models}
The first work addressing automated pronunciation assessment for the Holy Quran is RDI \cite{sherif2007enhancing}, which built a complete system for detecting pronunciation errors (Qalqala, Idgham, Iqlab) but omitted details on phoneticization. Subsequent work \cite{abdou2014computer, al2018computer} used DNNs to replace HMMs. Many studies rely on phoneme duration for duration-dependent rules like Madd and Ghunna \cite{mohammed2017recognition, alqadasi2023improving} but use limited datasets. Others focus on specific rules like Qalqala \cite{omran2023automatic} or Ghunna and Madd \cite{shaiakhmetov2025evaluation,10485145}. The work most aligned with our vision is \cite{putra2012developing}, which introduced a multi-level detection system (Makhraj and Tajweed) but relied on HMMs and minimal data.

While prior efforts focused on rule-based detection with constrained corpora, Tarteel \cite{khan2021tarteel} advanced the underlying ASR infrastructure by developing a robust system for diacritized character recognition, albeit without integrating Tajweed-specific error classification. Notably, they addressed the critical data bottleneck by releasing a crowd-sourced dataset of 25K utterances (68 hours) and later expanding this to 9K hours of private annotated data through application users.

A critical pattern across prior work is limitation in: (a) data scale (verse-restricted corpora), (b) rule coverage (one to three rules), and (c) open-source availability. By contrast, this work targets all Hafs Tajweed rules (except Ishmam) using a large expert-recitation corpus and releases both code and data.

\subsection{Pretrained Speech Encoders with Self-Supervised Learning (SSL)}
Speech pretraining began early \cite{hinton2006reducing} but was constrained by RNNs \cite{hopfield1982neural}. Transformers \cite{vaswani2017attention} enabled large-scale pretraining. BERT \cite{devlin2019bert} introduced masked language modeling, extended to speech with wav2vec \cite{schneider2019wav2vec} and wav2vec2.0 \cite{baevski2020wav2vec}. Conformer \cite{gulati2020conformer} integrated convolution. Google's Wav2Vec2-BERT \cite{chung2021w2v} applied MLM to speech. Facebook extended Wav2Vec2-BERT pretraining \cite{barrault2023seamless} to 4.5M hours (including 110K Arabic hours), ideal for low-resource fine-tuning.

Our work bridges the gap by pairing a large SSL encoder with QPS, a phonetic script that makes Tajweed structure explicit at the label level.

\section{Quran Phonetic Script}
\label{sec:qps}
We consider QPS to be the most valuable contribution of our work. By formalizing Holy Quran pronunciation assessment as an ASR problem represented through this script, we provide a comprehensive solution.

Modern Standard Arabic orthography cannot adequately represent Tajweed rules for error detection. Our phonetic script addresses this by capturing all Tajweed pronunciation errors except Ishmam (visual mouth movement without audible output). We based our script on classical Muslim scholarship rather than IPA for three reasons: (1) historical precedence (Muslim scholars from the 6th to 14th centuries defined Quranic errors centuries before modern phonetics), (2) scientific foundation (e.g., Al-Khalil ibn Ahmad systematically described articulations with remarkable accuracy \cite{article-khalil}), and (3) pedagogical relevance (learners' errors align with classical definitions).

Following \cite{sweed2021}, Quran recitation errors fall into three categories: Articulation Errors (incorrect phoneme exits), Attribute Errors (mistakes in Sifat al-Huruf), and Tajweed Rule Errors (incorrect application of rules). Our script addresses all three through two main output levels: Phonemes Level (letters, vowels, and Tajweed rules) and Sifat Level (10 articulation attributes). Refer to Tables~\ref{tab:phoneme_set} and \ref{tab:sifat_set} for the complete vocabularies.

Our script has some important characteristics: Normal Madd appears as consecutive madd symbols (e.g., 4-beat Madd: \arb{اااا}); Madd al-Leen is represented with multiple waw/yaa symbols; Stressed Ghunnah (e.g., \arb{النون المشددة}) is shown as three consecutive noon symbols (\arb{ننن}); Ikhfa is represented as three consecutive noon\_mokhfah (\arb{ںںں}) or meem\_mokhfah (\arb{۾۾۾}); \arb{إدغام بغنة} (Ghunnah assimilation) for the letters yaa and waw is represented by doubling (e.g., \arb{مَن يَعْمَلْ} $\rightarrow$ \arb{مَيييَعمَل}); Sakin is indicated by the absence of a following symbol; Imala is represented using fatha\_momala and alif\_momala; and Rawm is marked by a dama\_mokhtalasa marker.

Table~\ref{tab:examples_with_sifat} illustrates the phonetization process for the word (\arb{أَتُحَٰٓجُّوٓنِّى}), showing its conversion into phonetic components and Sifat attributes; refer to the table caption for a detailed row-by-row breakdown.

\begin{table*}[h]
	\centering
	\caption{Examples of Uthmani to Phonetic Script Conversion with Sifat Attributes. This example shows the phonetization of word (\arb{أَتُحَٰٓجُّوٓنِّى}) row by row: The first row shows conversion of (\arb{أَ}) to (\arb{ءَ}) with its sifat in the row, following the second row. For the fourth row, showing the madd Lazim rule with 6 beats phoneticized as 6 alifs (\arb{اااااا}), same as the sixth row but with damma represented as 6 waw\_madd (\arb{ۥۥۥۥۥۥ}). And for the fifth row we notice that we converted (\arb{جُّ}) to (\arb{ججُ}) disassembling shadda. The last row shows the normal madd of yaa with two beats represented as two yadd\_madd (\arb{ۦۦ})}
	\label{tab:examples_with_sifat}
	\scriptsize
	\setlength{\tabcolsep}{3pt}
	\begin{tabular}{@{}>{\centering\arraybackslash}m{1.2cm} >{\centering\arraybackslash}m{1.5cm} *{10}{>{\centering\arraybackslash}m{0.7cm}@{}}}
		\toprule
		\textbf{Uthmani} & \textbf{Phonetic} &
		\textbf{H/J}     &
		\textbf{S/R}     &
		\textbf{T/T}     &
		\textbf{Itb}     &
		\textbf{Saf}     &
		\textbf{Qal}     &
		\textbf{Tik}     &
		\textbf{Taf}     &
		\textbf{Ist}     &
		\textbf{Gho}                                                                                     \\
		\cmidrule(lr){1-1} \cmidrule(lr){2-2} \cmidrule(lr){3-12}
		\arb{أَ}          & \arb{ءَ}           & jahr & shd & mrq & mnf & no & nql & nkr & ntf & nst & nmg \\
		\arb{تُ}          & \arb{تُ}           & hams & shd & mrq & mnf & no & nql & nkr & ntf & nst & nmg \\
		\arb{حَـ}         & \arb{حَ}           & hams & rkh & mrq & mnf & no & nql & nkr & ntf & nst & nmg \\
		\arb{ـٰٓ}          & \arb{اااااا}      & hams & rkh & mrq & mnf & no & nql & nkr & ntf & nst & nmg \\
		\arb{جُّ}          & \arb{ججُ}          & jahr & shd & mrq & mnf & no & nql & nkr & ntf & nst & nmg \\
		\arb{وٓ}          & \arb{ۥۥۥۥۥۥ}      & jahr & rkh & mrq & mnf & no & nql & nkr & ntf & nst & nmg \\
		\arb{نِّ}          & \arb{ننننِ}        & jahr & btw & mrq & mnf & no & nql & nkr & ntf & nst & mg  \\
		\arb{ى}          & \arb{ۦۦ}          & jahr & rkh & mrq & mnf & no & nql & nkr & ntf & nst & nmg \\
		\bottomrule
	\end{tabular}

	\vspace{2mm}
	\begin{center}  
		\scriptsize
		\textbf{Attribute Abbreviations:} \\
		H/J: Hams/Jahr \quad S/R: Shidda/Rakhawa \quad T/T: Tafkheem/Taqeeq \quad Itb: Itbaq \\
		Saf: Safeer \quad Qal: Qalqla \quad Tik: Tikraar \quad Taf: Tafashie \quad Ist: Istitala \quad Gho: Ghonna \\

		\textbf{Value Abbreviations:} \\
		shd: shadeed \quad rkh: rikhw \quad btw: between \quad mrq: moraqaq \\
		mof: mofakham \quad mnf: monfateh \quad mtb: motbaq \quad no: no\_safeer \\
		nql: not\_moqalqal \quad nkr: not\_mokarar \quad ntf: not\_motafashie \\
		nst: not\_mostateel \quad nmg: not\_maghnoon \quad mg: maghnoon
	\end{center}  
\end{table*}

\subsection{Development Methodology}
Our phonetization has two steps:
\begin{enumerate}
	\item \textbf{Imlaey to Uthmani Conversion}: We selected Uthmani script as our foundation because it contains specialized Tajweed diacritics (Madd, Tasheel, etc.) and preserves pause rules critical for recitation (e.g., stopping on \arb{رحمت}). To do so, we created an annotation UI to manually annotate misaligned words in both scripts (e.g., \ref{tab:imlaey_to_uthmani_ex}), and then developed an algorithm that relies on these annotations to convert Imlaey to Uthmani.
	\item \textbf{Uthmani to Phonetic Script Conversion} \\
	      We implemented the process through 26 sequential operations. Each operation contains one or more regular expressions, as shown in the Appendix~\ref{subsec:conversion_ops}.

	\item \textbf{Extracting Sifat}: Next, we extract the 10 attributes (Sifat) defined in Table~\ref{tab:sifat_set}, excluding \textbf{Inhiraf} (\arb{إنحراف}), as it describes the shidda/rakhawa spectrum, and \textbf{Leen} (\arb{اللين}), as it was already handled through our Madd representation.
\end{enumerate}

\begin{table}[t]
	\caption{Example of misalignment between Uthmani and Imlaey Scripts}
	\label{tab:imlaey_to_uthmani_ex}
	\centering
	\begin{tabular}{|c|c|}
		\hline
		\textbf{Imlaey Script} & \textbf{Uthmani Script} \\
		\hline
		\arb{يَا ابْنَ أُمَّ}        & \arb{يَبْنَؤُمَّ}             \\
		\hline
	\end{tabular}
\end{table}

\section{Data Preparation}
To prepare the data, we first defined selection criteria. We aimed to collect recitations from the best reciters worldwide to serve as references for judging Quran learners. In our study, we considered only \textit{Hafs} riwayah (\arb{رواية حفص}) as it's the most popular recitation method globally. Recognizing that manual data annotation requires significant effort and time, we created a 98\% automated pipeline as for data collection shown in Figure~\ref{fig:data_pipeline}. The steps are:
\begin{enumerate}
    \item \textbf{Digitized Quran:} We chose Tanzil (standard Unicode, both Imlaei and Uthmani versions) over KFGQPC due to stability.
    \item \textbf{Variant Criteria for Hafs:} We defined variants (e.g., Madd al-Munfasil: 2,4,5, or 6 beats) through Qira'at literature \cite{al-dabbaa}.
    \item \textbf{Expert Recitation Collection:} Criteria: perfect recitation, studio recording, complete Mushaf, high audio quality.
    \item \textbf{Segmentation:} We fine-tuned Wav2Vec2-BERT \cite{barrault2023seamless} for frame-level classification. Table~\ref{tab:seg_results} shows results on unseen Mushafs.
    \item \textbf{Transcription:} We used Tarteel ASR \cite{tarteel_whisper_ar_quran} (Whisper fine-tuned on Quranic recitations \cite{radford2023robust}) with 5.75\% WER.
    \item \textbf{Verification (Tasmeea Algorithm):} A fuzzy matching algorithm based on Levenshtein distance \cite{levenshtein1966binary} corrects transcription errors. Samples below 85\% threshold are sent to human annotators.
\end{enumerate}

We define a \textit{Moshaf} as a complete Quran recitation (chapters 1-114) by a specific reciter. Statistics are summarized in table~\ref{tab:moshaf}. We manually annotated 5400 samples out of 286,537 utterances, resulting for the automation ratio of 98\%.

\begin{table}[t]
	\centering
	\caption{Dataset Statistics per Moshaf}
	\label{tab:moshaf}
	\begin{tabular}{ccc}
		\hline
		\textbf{Moshaf ID} & \textbf{Hours}       & \textbf{Length} \\
		\hline
		0.0                & 28.48                & 9133            \\
		0.1                & 40.31                & 10764           \\
		0.2                & 49.47                & 9971            \\
		0.3                & 37.19                & 12604           \\
		1.0                & 28.41                & 10939           \\
		2.0                & 51.05                & 9942            \\
		2.1                & 30.03                & 10394           \\
		3.0                & 25.19                & 10444           \\
		4.0                & 29.12                & 10994           \\
		5.0                & 28.02                & 11482           \\
		6.0                & 39.39                & 12435           \\
		7.0                & 28.26                & 9907            \\
		8.0                & 30.86                & 10330           \\
		9.0                & 27.95                & 10642           \\
		11.0               & 24.01                & 10363           \\
		12.0               & 33.42                & 9880            \\
		13.0               & 33.99                & 9377            \\
		19.0               & 30.11                & 11278           \\
		22.0               & 28.11                & 10332           \\
		24.0               & 28.51                & 9868            \\
		25.0               & 16.93                & 7922            \\
		26.0               & 30.44                & 11565           \\
		26.1               & 32.71                & 11850           \\
		27.0               & 28.05                & 11213           \\
		28.0               & 31.05                & 10535           \\
		29.0               & 27.79                & 11061           \\
		30.0               & 29.14                & 11312           \\
		\hline
		\textbf{Total}     & \textbf{847.9944402} & \textbf{286537} \\
		\hline
	\end{tabular}
\end{table}

\subsection{Choose a Digitized Version of the Holy Quran}
The Quran has multiple digitized versions including Tanzil\footnote{\url{https://tanzil.net}} and King Fahd Complex\footnote{\url{https://qurancomplex.gov.sa}}. We chose Tanzil because:
\begin{itemize}
	\item It uses standard Unicode characters
	\item Contains both \textit{Imlaei} and \textit{Uthmani} versions
	\item Maintains high accuracy
\end{itemize}
We excluded KFGQPC due to its evolving/unstable nature compared to Tanzil.

\subsection{Define Variant Criteria for Hafs}
\textit{Hafs} riwayah contains variants, e.g., \textit{Madd Al-Munfasil} (\arb{مد المنفصل}) can extend 2, 4, 5, or 6 beats. We rigorously defined these variants through the Qira'at literature \cite{al-dabbaa}, summarized in the following attributes in the Appendix section \ref{sec:moshaf_attributes}.

\subsection{Collect Expert Recitations}
We collected recitations from 22 world-class reciters with premium audio quality, totaling \textbf{893 hours} pre-filtering.

To ensure high-quality reference data for judging Quran learners, reciters were selected based on the following criteria: perfect recitation with no mistakes, non-prayer (studio) recording not during prayer, complete Moshaf (full Quran, chapters 1-114), and high audio quality. All recitations were recorded in high-quality studio conditions with thorough review to ensure high quality.

\begin{table}[t]
	\centering
	\caption{Distribution of Reciters by Country}
	\label{tab:reciters_country}
	\begin{tabular}{lc}
		\hline
		\textbf{Country Name}     & \textbf{Number of Reciters} \\
		\hline
		Egypt (EG)                & 15                          \\
		Saudi Arabia (SA)         & 3                           \\
		Syria (SY)                & 1                           \\
		Somalia (SO)              & 1                           \\
		Kuwait (KW)               & 1                           \\
		United Arab Emirates (AE) & 1                           \\
		\hline
		\textbf{Total}            & \textbf{22}                 \\
		\hline
	\end{tabular}
\end{table}

\begin{figure}[t]
	\centering
	\includegraphics[width=0.8\linewidth]{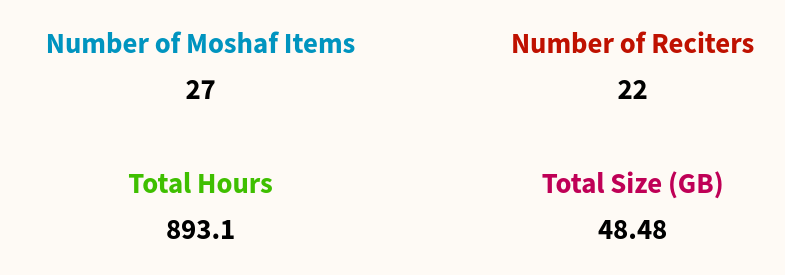}
	\caption{Database Collection Statistics}
	\label{fig:stats}
\end{figure}

\begin{figure*}[!htbp]
	\centering
	\includegraphics[width=0.8\linewidth]{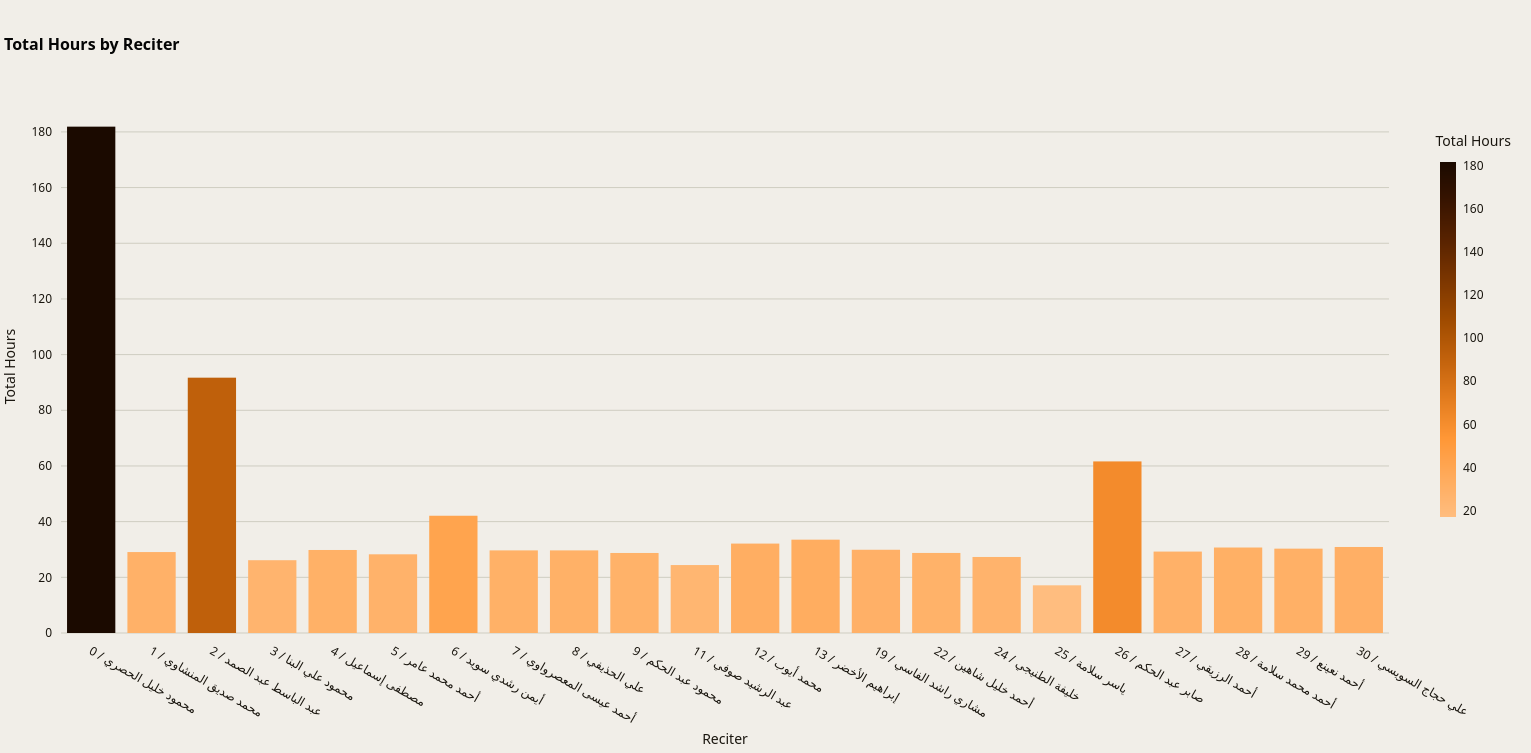}
	\caption{Reciters Statistics}
	\label{fig:reciters}
\end{figure*}

We developed a web GUI using Streamlit that downloads and extracts metadata for each track, organizes data by Moshaf (each chapter as "001.mp3"), and annotates Moshaf attributes.

\subsection{Segment Recitations}
Since Tajweed rules are affected by pauses (\arb{وقف}), accurate segmentation is crucial. We initially tested open-source Voice Activity Detection (VAD) models including SileroVAD \cite{SileroVAD} and PyAnnotate \cite{Plaquet23}. Poor Quran-specific performance led us to develop a custom segmenter by fine-tuning Wav2Vec2-BERT\footnote{\url{https://github.com/obadx/recitations-segmenter}} \cite{barrault2023seamless} for frame-level classification.

\paragraph{Preparing Segmenter Data} We selected mosahf compatible with SileroVAD v4, using EveryAyah\footnote{\url{https://everyayah.com/}} (pre-segmented by ayah) as ground truth. After tuning parameters per Moshaf, we included: threshold, minimum silence duration (to merge segments), minimum speech duration (to discard short segments), and padding (added at segment boundaries).

\paragraph{Data Augmentation} Using the Audiomentations \cite{Audiomentations} library, we replicated SileroVAD's noise setup on 40\% of samples, adding \texttt{TimeStretch} (0.8x-1.5x) to simulate recitation speeds and sliding window truncation (1-second windows) for long samples instead of exclusion.

\paragraph{Training Segmenter}
We fine-tuned Wav2Vec2-BERT for frame classification for 1 epoch using the AdamW optimizer with a constant learning rate of 5e-5, default betas and bfloat16 precision on an Nvidia L40 GPU, with a batch size of 50 and a maximum speech sample length of 20 seconds.


\begin{figure*}[!htbp]
	\centering
	\includegraphics[width=0.85\textwidth]{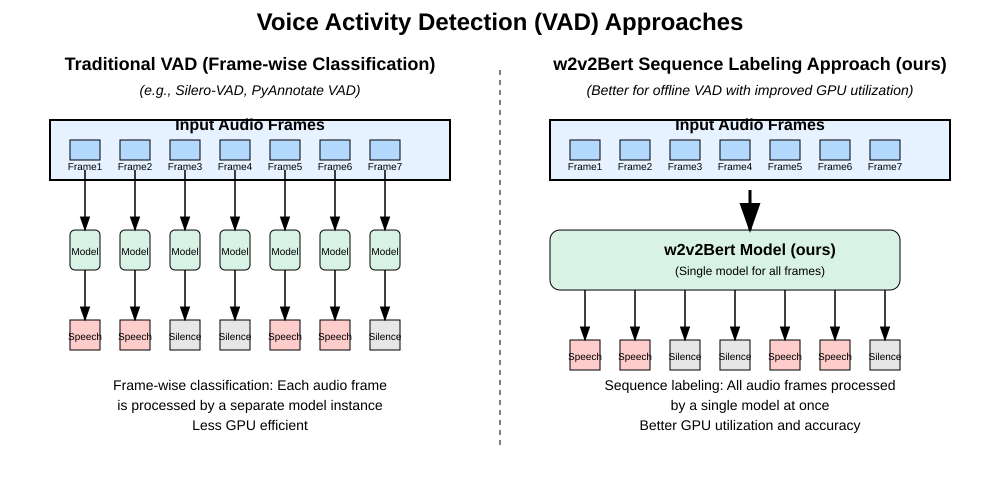}
	\caption{VAD architecture vs. standard streaming models}
	\label{fig:vad-arch}
\end{figure*}

Results of our segmenter on unseen mosahf in table~\ref{tab:seg_results}:

\begin{table}[t]
	\centering
	\caption{Test results of the segmenter on unseen full moshaf. The result is validated by actual usage of the segmenter}
	\label{tab:seg_results}
	\vspace{2pt}
	\begin{tabular}{lc}
		\hline
		\textbf{Metric} & \textbf{Value} \\
		\hline
		Test Loss       & 0.0277         \\
		Test Accuracy   & 0.9935         \\
		Test F1 Score   & 0.99476        \\
		\hline
	\end{tabular}
\end{table}

\subsection{Transcribe Segmented Parts}
We employed Tarteel ASR \cite{tarteel_whisper_ar_quran} (Whisper fine-tuned on Quranic recitations \cite{radford2023robust}) with a WER of 5.7544\%. To handle its 30-second limit, we used sliding window truncation (10-second windows), with verification in the next step.

\subsection{Automated Data Pipeline Assessment}
In this section, we discuss how we assessed the pipeline performance phase by phase:

\textbf{Segmentation Verification}: For the segmentation phase, we performed manual inspection of 50--75 random samples per Moshaf, in addition to short ($< 3$ seconds) and long ($> 35$ seconds) tracks. Any Moshaf with accuracy $< 99\%$ was excluded; specifically, Moshaf 25.0 was excluded due to poor segmentation ($90\%$).

\textbf{Transcription Verification}: We utilized the Tarteel model, which has a 5.7544\% Word Error Rate (WER). To correct potential transcription errors, we leveraged the fact that our input consists of 100\% correct expert recitations to design the \textit{Tasmeea}-inspired algorithm. This algorithm matches the Tarteel transcription to the Quranic text—analogous to a student reciting to a teacher—using fuzzy matching based on Levenshtein distance \cite{levenshtein1966binary} with a moving pointer. If the text matching threshold falls below $85\%$, the samples are referred to human annotators for correction.

The result is a high-quality dataset with no missing verses across 27 full Quran recitations. Refer to the Tasmeea Algorithm in the Appendix \ref{alg:tasmeea}

\begin{figure*}[!htbp]
	\centering
	\begin{tikzpicture}[node distance=2cm, auto]
		\tikzstyle{process} = [rectangle, minimum width=3cm, minimum height=1cm, text centered, draw=blue!60, fill=blue!10, rounded corners]
		\tikzstyle{data} = [rectangle, minimum width=2.5cm, minimum height=1cm, text centered, draw=green!60, fill=green!10, rounded corners]
		\tikzstyle{arrow} = [thick,->,>=stealth,draw=gray!70]

		\node[data] (collection) {Expert Recitation Collection};
		\node[process, below of=collection] (segmentation) {Segmentation at Waqf Points};
		\node[process, below of=segmentation] (transcription) {ASR Transcription};
		\node[process, below of=transcription] (verification) {Tasmeea Verification};
		\node[process, below of=verification] (encoding) {QPS Encoding};
		\node[data, below of=encoding] (dataset) {Final Dataset};

		\draw[arrow] (collection) -- (segmentation);
		\draw[arrow] (segmentation) -- (transcription);
		\draw[arrow] (transcription) -- (verification);
		\draw[arrow] (verification) -- (encoding);
		\draw[arrow] (encoding) -- (dataset);

		\node[right of=collection, xshift=3cm, text width=4cm, text=gray] {\small 22 world-class reciters};
		\node[right of=segmentation, xshift=3cm, text width=4cm, text=gray] {\small wav2vec2-BERT model};
		\node[right of=transcription, xshift=3cm, text width=4cm, text=gray] {\small Tarteel Whisper};
		\node[right of=verification, xshift=3cm, text width=4cm, text=gray] {\small Validation algorithm};
		\node[right of=encoding, xshift=3cm, text width=4cm, text=gray] {\small 11-level script};
		\node[right of=dataset, xshift=3cm, text width=4cm, text=gray] {\small 848 hours, 286k utterances};
	\end{tikzpicture}
	\caption{The complete data preparation pipeline showing the six main stages from expert recitation collection to the final dataset generation. Each stage utilizes specialized models and algorithms to ensure high-quality annotations.}
	\label{fig:data_pipeline}
\end{figure*}

\section{Modeling The Quran Phonetic Script}
Our Quran Phonetic script has two outputs: \texttt{phonemes} and \texttt{sifat} (which has 10 attributes). We implemented this as a speech encoder $E_{\theta}(x)$ followed by 11 linear heads $W_{k}$, representing 11 parallel transcription levels (\texttt{phonemes} and the 10 \texttt{sifat} attributes). We chose CTC loss \cite{graves2006ctc} without language model integration because we aim to capture what the user actually said, not what they intended to say. We name our architecture \textbf{Multi-level CTC}. For each level $k$, the CTC loss is $L_{k}$, and the final weighted loss is:
\begin{equation}
	L = \sum_{k=1}^{11} w_{k} L_{k}, \quad \text{subject to} \quad \sum_{k=1}^{11} w_{k} = 1
\end{equation}

The weights are assigned based on the vocabulary size of each level to balance their contribution. Specifically, the \texttt{phonemes} level (vocabulary size 44, including blank token) receives a weight of 0.4; the \texttt{shidda\_or\_rakhawa} and \texttt{tafkheem\_or\_taqeeq} levels (each with a vocabulary size of 4 including blank token) receive a weight of 0.0605 each. The remaining eight levels (each with a vocabulary size of 3 including blank token) receive a weight of 0.069875 each.

\begin{figure*}[!htbp]
	\centering
	\includegraphics[width=\linewidth]{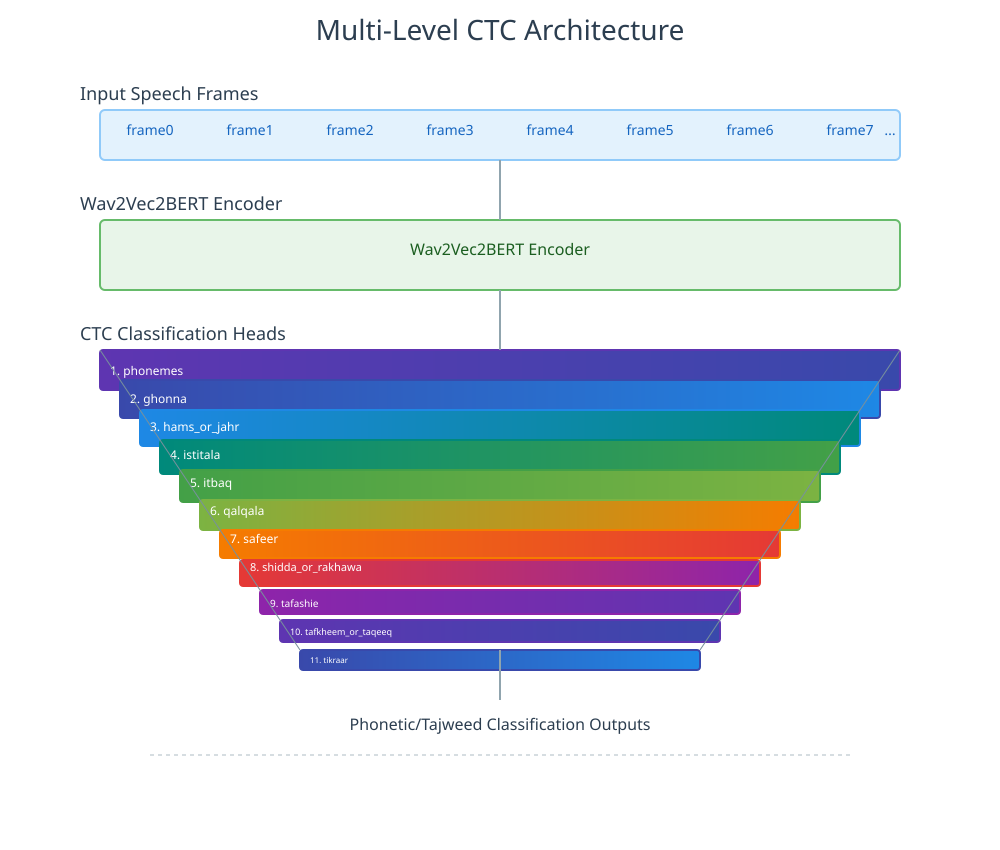}
	\caption{Multi-level CTC loss Architecture composed of 11 Heads for every level and CTC loss for every level with weighted average loss}
	\label{fig:multilevel-ctc}
\end{figure*}

We fine-tuned Facebook's Wav2Vec2-Bert \cite{barrault2023seamless} for a single epoch using an adapter layer and the AdamW optimizer with its default $\beta$ values. We employed a constant learning rate of \texttt{5e-5}, a batch size of 64, and a maximum sample length of 30 seconds, with training performed in \texttt{bfloat16} precision. We applied augmentations identical to Silero VAD \cite{SileroVAD} using the \texttt{audiomentations} library \cite{Audiomentations}, with additional augmentations: \texttt{TimeStretch} and \texttt{GainTransition}. We filtered out samples longer than \texttt{30 seconds} for efficient GPU utilization, sacrificing only 3k samples out of 250k training samples. The training was conducted on an H200 GPU with 141 GB of GPU memory for 7 hours.

Due to our limited budget, we conducted preliminary experiments to determine the optimal hyperparameters by training on a subset of the training data (4 Mushafs) on Google Colab. We evaluated different training setups, including linear and cosine schedulers, and the impact of SpecAugment \cite{park2019specaugment}. The best results were achieved with a constant learning rate and without SpecAugment, as the latter significantly degraded performance. Based on these findings, we proceeded with the full training on a high-performance GPU.

\begin{figure}[t]
	\centering
	\includegraphics[width=\linewidth]{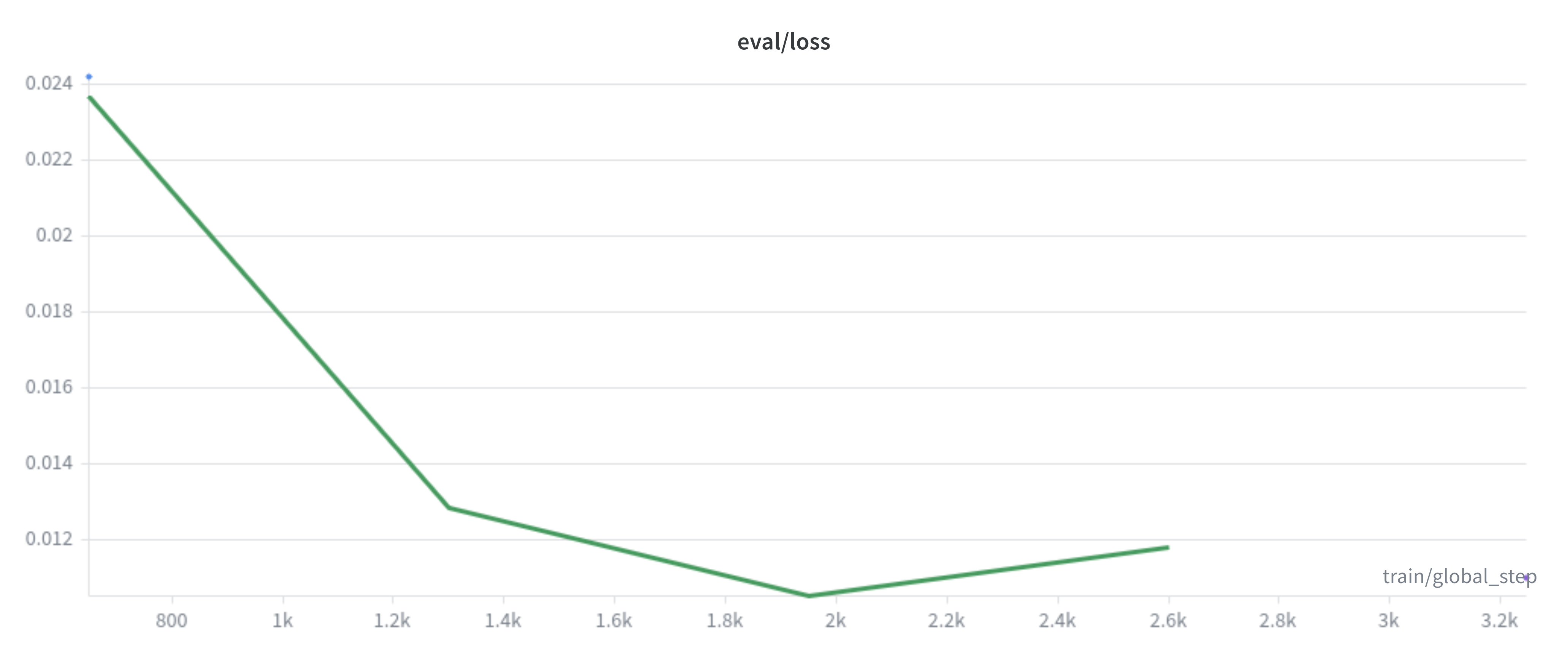}
	\caption{Evolution of the evaluation loss during training.}
	\label{fig:eval_loss}
\end{figure}

The convergence behavior of the model is shown in Figure \ref{fig:eval_loss}, where the evaluation loss decreases consistently throughout the training process, confirming the stability of the multi-level CTC objective.

\begin{figure}[t]
	\centering
	\includegraphics[width=\linewidth]{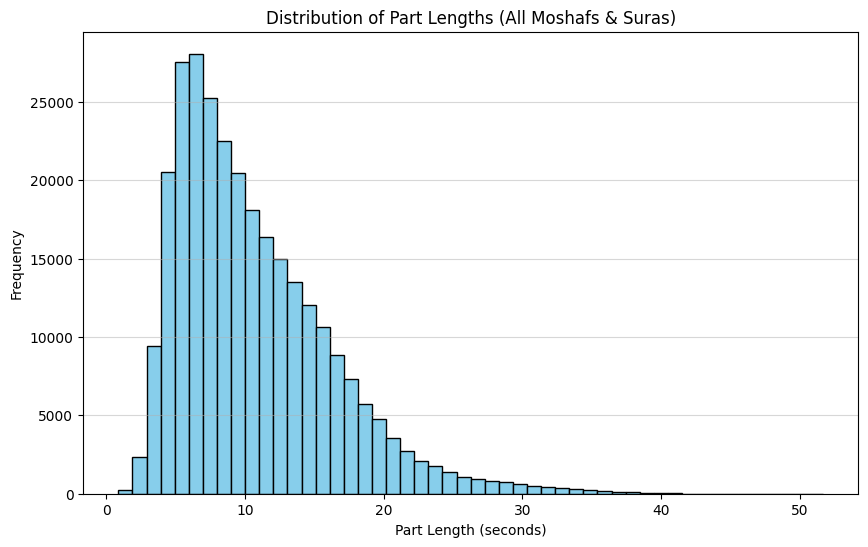}
	\caption{Recitations lengths in seconds for the whole dataset}
	\label{fig:audio-lengths}
\end{figure}

\section{Results}
\subsection{Expert Recitation Evaluation}
We trained on all available Moshaf datasets, reserving three Moshaf (19.0, 29.0, 30.0) for comprehensive testing. These test datasets feature expert male reciters with extensive training in Tajweed rules. The expert nature of these recordings provides an ideal evaluation environment for assessing the model's fundamental phonetic transcription capabilities across different representation levels. Notably, the phonemes level presents the greatest challenge with a 44-character vocabulary (including padding), resulting in the highest Phoneme Error Rate of 0.543\% and average PER of 0.21\% across all levels as shown in Table~\ref{tab:results}, while still demonstrating excellent overall performance that validates our multi-level CTC approach.

\begin{table}[t]
	\centering
	\caption{Test Results on Expert Quranic Recitations on three Moshaf datasets (19.0, 29.0, 30.0). The phonemes level has the highest PER (0.543\%) due to its 44-character vocabulary including padding.}
	\vspace{3pt}
	\label{tab:results}
	\begin{tabular}{lc}
		\hline
		\textbf{Metric}           & \textbf{Value}  \\
		\hline
		loss                      & 0.01162         \\
		per\_phonemes             & 0.00543         \\
		per\_hams\_or\_jahr       & 0.00117         \\
		per\_shidda\_or\_rakhawa  & 0.00172         \\
		per\_tafkheem\_or\_taqeeq & 0.00167         \\
		per\_itbaq                & 0.00092         \\
		per\_safeer               & 0.00132         \\
		per\_qalqla               & 0.00085         \\
		per\_tikraar              & 0.0009          \\
		per\_tafashie             & 0.0016          \\
		per\_istitala             & 0.0008          \\
		per\_ghonna               & 0.0013          \\
		average\_per              & \textbf{0.0021} \\
		\hline
	\end{tabular}
\end{table}

\subsection{Learner Error Evaluation (qdat\_bench)}
To evaluate our model's performance on real recitation errors, we tested on our developed qdat\_bench benchmark, which enhances the original \cite{osman2021qdat} dataset through extensive expert reannotation and the addition of all Tajweed rules, including phonemes and 10 sifat levels. We performed bootstrap analysis with 10,000 iterations to provide comprehensive view of metric distributions. The results, shown in Table~\ref{tab:qdat_bench}, demonstrate the model's effectiveness in detecting pronunciation errors in authentic learner recordings.

\begin{table*}[t]
	\centering
	\caption{qdat\_bench Results on Authentic Learner Recordings. Evaluated on 159 samples (120 female, 39 male) from Surah Al-Ma'idah (5:109). F1 for Tajweed rules: Noon Moshaddadah, Ikhfaa, Qalqlah. RMSE for Madd lengths: normal, separated, and aared madd.}
	\vspace{3pt}
	\label{tab:qdat_bench}
	\begin{tabular}{lcc}
		\hline
		\textbf{Aggregate Metrics} & \textbf{Original} & \textbf{Bootstrap (n=10,000)} \\
		\hline
		per\_phonemes              & 0.0578            & 0.0603 $\pm$ 0.0053           \\
		avg\_per                   & 0.0194            & 0.0197 $\pm$ 0.0032           \\
		avg\_tajweed\_f1           & 0.7582            & 0.7575 $\pm$ 0.0177           \\
		avg\_tajweed\_acc          & 0.8470            & 0.8464 $\pm$ 0.0174           \\
		avg\_madd\_rmse            & 0.5965            & 0.5960 $\pm$ 0.0374           \\
		avg\_normal\_madd\_rmse    & 0.4645            & 0.4688 $\pm$ 0.0501           \\
		rmse\_madd\_aared          & 1.0340            & 1.0251 $\pm$ 0.0724           \\
		rmse\_separate\_madd       & 0.6758            & 0.6868 $\pm$ 0.0582           \\
		\hline
		\textbf{Tajweed Rules F1}  & \textbf{Original} & \textbf{Bootstrap (n=10,000)} \\
		\hline
		Noon Moshaddadah           & 0.8691            & 0.8682 $\pm$ 0.0324           \\
		Ikhfaa (Noon Mokhfah)      & 0.4526            & 0.4510 $\pm$ 0.0259           \\
		Qalqalah                   & 0.9530            & 0.9532 $\pm$ 0.0174           \\
		\hline
		\textbf{Madd Rules RMSE}   & \textbf{Original} & \textbf{Bootstrap (n=10,000)} \\
		\hline
		Normal Madd (5 rules)      & 0.4645            & 0.4688 $\pm$ 0.0501           \\
		Separate Madd              & 0.6868            & 0.6758 $\pm$ 0.0582           \\
		Aared Madd                 & 1.0340            & 1.0251 $\pm$ 0.0724           \\
		\hline
	\end{tabular}
\end{table*}

qdat\_bench shows a higher PER of 0.058 on authentic learner recordings, which is expected given the complexity of detecting real pronunciation errors and handling variability in learner execution. However, 5.8\% remains very acceptable despite being trained on expert recitations only. The aggregate Tajweed F1 score of 0.758 represents the mean across three key rules: Noon Moshaddadah (0.869), Ikhfaa (0.453), and Qalqalah (0.953). Similarly, the average Madd RMSE of 0.596 encompasses normal madd (0.464), separate madd (0.687), and aared madd (1.034).

\begin{figure*}[!htbp]
	\centering
	\includegraphics[width=0.9\textwidth]{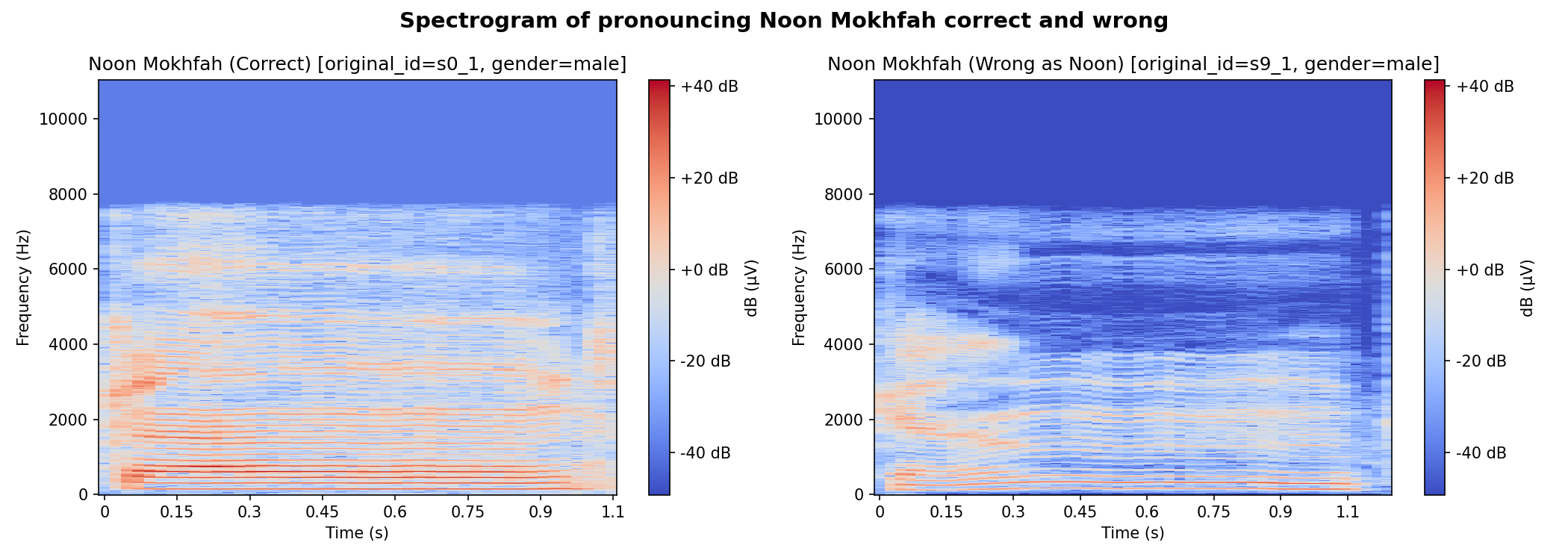}
	\caption{The figure illustrates the challenge of distinguishing Ikhfaa pronunciation from clear noon pronunciation for both humans and AI models. Spectrogram comparison of Noon Mokhfah (Ikhfaa) pronunciation from qdat\_bench: (left) correct Ikhfaa pronunciation of \arb{أنت} by a male reciter (original\_id=s0\_1), and (right) incorrect elongated noon pronunciation by a male reciter (original\_id=s9\_0) that the model failed to detect as a pronunciation error.}
	\label{fig:noon_mokhfah_spectrogram}
\end{figure*}

Notably, the Ikhfaa F1 score of 0.453 is considerably lower than other rules, which is expected given the acoustic similarity between Ikhfaa and clear noon pronunciation (see Figure~\ref{fig:noon_mokhfah_spectrogram}). This challenge affects both human perception and automated detection. The model's performance on this rule reflects the inherent difficulty in detecting subtle nasalization differences. Detailed analysis is provided in the appendix~\ref{sec:qdat_bench}.

Notably, despite being trained exclusively on male expert reciters, our model demonstrates promising generalization by achieving 84.7\% accuracy on female recitations in qdat\_bench, highlighting robustness for real-world deployment. This performance validates the practical applicability of our Quranic phonetic script and multi-level CTC architecture.

\begin{figure*}[!htbp]
	\centering
	\includegraphics[width=1\linewidth]{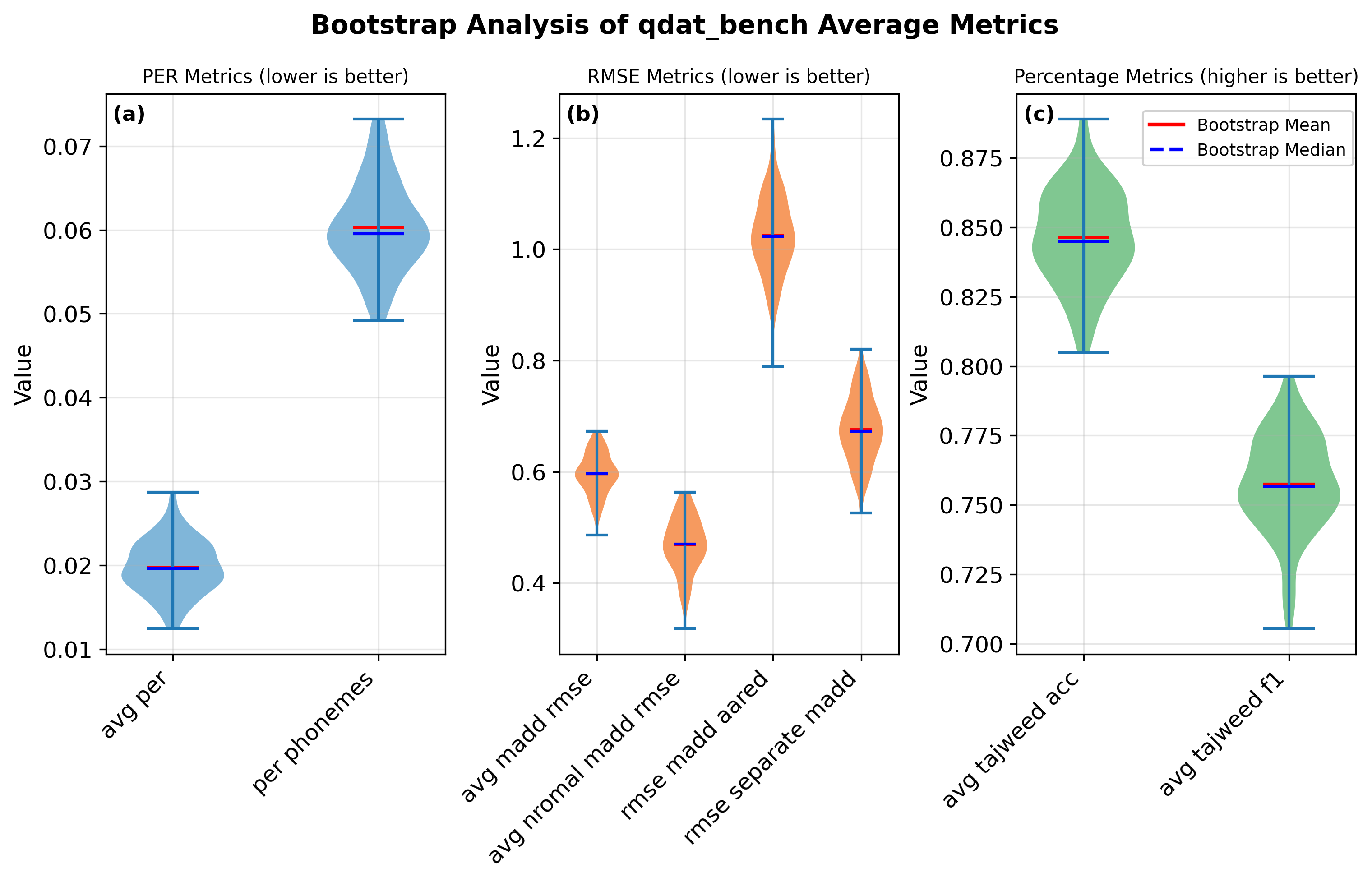}
	\caption{Violin plot for bootstrap analysis (n=10,000) of qdat\_bench aggregate metrics. (a) Phoneme Error Rate (PER) metrics (lower is better), (b) RMSE metrics for Madd rules (lower is better), and (c) Tajweed performance metrics (higher is better). Dashed lines indicate original metric values before bootstrapping.}
	\label{fig:qdat_bootstrap_violin_results}
\end{figure*}

Figure~\ref{fig:qdat_bootstrap_violin_results} visualizes the bootstrap distributions for aggregate metrics, showing stable estimates with narrow confidence intervals.

Despite these promising results, certain model limitations should be acknowledged. Attribute-specific articulation patterns present inherent challenges: certain phonetic attributes apply exclusively to individual letters, such as \texttt{Istitala} for (\arb{ض}) and \texttt{Tikrar} for (\arb{ر}). Future work should focus on annotating real data with these specific errors to improve model understanding (see Section~\ref{sec:limitations} for detailed discussion).

\begin{figure*}[!htbp]
	\centering
	\includegraphics[width=0.9\textwidth]{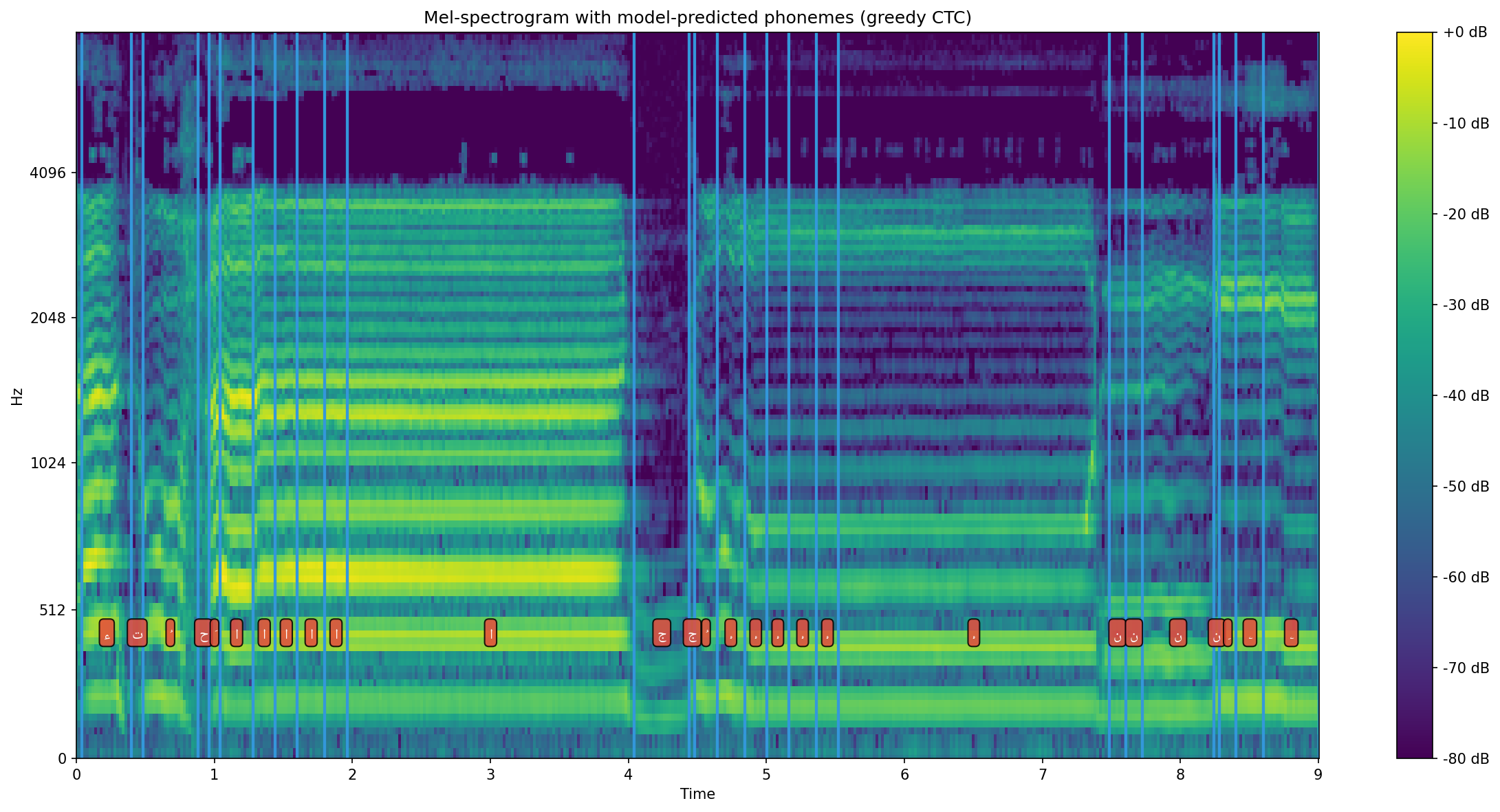}
	\caption{Spectrogram analysis of the aligned model output for the word \arb{أَتُحَٰٓجُّوٓنِّى} (refer to Table~\ref{tab:examples_with_sifat} for the complete reference phonetization). Reference: \arb{ءَتُحَاااااااججُۥۥۥۥۥۥننننِۦۦ}, Predicted: \arb{ءتُحَاااااججُۥۥۥۥۥۥننننِۦۦ}. The phoneme boundaries remain near perfect; however, one error is observed: the fatha on the initial hamza (\arb{ءَ} → \arb{ء}) was missed. Additionally, the Madd alif (\arb{اااااا}) and Madd waw (\arb{ۥۥۥۥۥۥ}) durations are not balanced, with the final beat of the Madd sequence being notably larger.}
	\label{fig:tohajoni_aligned_output}
\end{figure*}

\section{Limitations and Future Work}
\label{sec:limitations}
A major limitation appears from attribute-specific articulation patterns: Certain attributes apply exclusively to individual letters, such as \texttt{Istitala} for (\arb{ض}) and \texttt{Tikrar} for (\arb{ر}). Consequently, we expect our model will be unable to capture instances of (\arb{ض}) without \texttt{Istitala} or (\arb{ر}) without \texttt{Tikrar}. This limitation similarly applies to Tajweed rules that occur less frequently in the Holy Quran, such as \texttt{Imala}, \texttt{Rawm}, and \texttt{Tasheel}. The solution is to annotate real data with these errors.

Additionally, the training dataset comprises only expert male reciters, which may limit generalizability to diverse vocal characteristics. The current scope is restricted to Hafs recitation style; future work will extend to other Riwayah (e.g., Warsh). Finally, qdat\_bench uses a single annotator and one verse; future versions should include multiple experts and broader Quranic coverage.


\section{Conclusion}
We present a novel approach for assessing pronunciation errors in Holy Quran learners through a multi-level Quran Phonetic Script that captures all pronunciation errors for \textit{Hafs} (except \texttt{Ishmam}, as it is a visual diacritic not orally produced). We provide 848 hours of annotated audio data with 286K samples, a 98\% automated pipeline for generating similar datasets featuring our Tasmeea verification algorithm, and a novel multi-level CTC model with an 11-level structure (1 phoneme level + 10 sifat levels). Achieving a 0.21\% average phoneme error rate on unseen expert test data proves the learnability of the Quran Phonetic Script. Furthermore, our model demonstrates robust generalization on real learner errors through qdat\_bench, achieving 75.8\% Tajweed F1 score and 84.7\% accuracy on female reciters, fundamentally transforming Holy Quran pronunciation assessment methodology.

\section*{Acknowledgement}
We express our profound gratitude to several individuals and organizations: Azza Abdolmiem, for her unwavering support and for providing a peaceful environment throughout the research period; Sheikh Ahmed Abdelsalam, Sheikh Mustafa Fathy, and Sheikh Mohamed Rabee, for their invaluable guidance in understanding and representing Tajweed rules and learner mistakes. We also thank BA-HPC\footnote{\url{https://hpc.bibalex.org/}} for providing access to high-performance computing resources and facilitating data processing. Special appreciation goes to Engineer Khaled Bahaa for his assistance with the payment method for GPUs.

\bibliography{references}
\bibliographystyle{icml2026}

\appendix
\onecolumn
\section{Appendix}\label{Appendix}
\subsection{Quran Phoneme Script Vocabulary}

\begin{table*}[t]
	\centering
	\caption{Phoneme Set (43 Symbols)}
	\label{tab:phoneme_set}
	\begin{tabular}{>{\raggedright}p{0.7\linewidth}c}
		\toprule
		\textbf{Phoneme Name} & \textbf{Symbol} \\
		\midrule
		hamza & \arb{ء} \\
		baa & \arb{ب} \\
		taa & \arb{ت} \\
		thaa & \arb{ث} \\
		jeem & \arb{ج}  \\
		haa\_mohmala & \arb{ح}  \\
		khaa & \arb{خ} \\
		daal & \arb{د}  \\
		thaal & \arb{ذ}  \\
		raa & \arb{ر}  \\
		zay & \arb{ز}  \\
		seen & \arb{س}  \\
		sheen & \arb{ش}  \\
		saad & \arb{ص}  \\
		daad & \arb{ض}  \\
		taa\_mofakhama  & \arb{ط} \\
		zaa\_mofakhama  & \arb{ظ} \\
		ayn & \arb{ع}  \\
		ghyn & \arb{غ}  \\
		faa & \arb{ف}  \\
		qaf & \arb{ق}  \\
		kaf & \arb{ك}  \\
		lam & \arb{ل}  \\
		meem & \arb{م}  \\
		noon & \arb{ن}  \\
		haa & \arb{ه}  \\
		waw & \arb{و}  \\
		yaa & \arb{ي}  \\
		alif & \arb{ا}  \\
		yaa\_madd & \arb{ۦ} \\
		waw\_madd & \arb{ۥ} \\
		fatha & \arb{َ}   \\
		dama & \arb{ُ}    \\
		kasra & \arb{ِ}  \\
		fatha\_momala & \arb{۪}  \\
		alif\_momala & \arb{ـ}  \\
		hamza\_mosahala & \arb{ٲ}   \\
		qlqla & \arb{ڇ}   \\
		noon\_mokhfah & \arb{ں}  \\
		meem\_mokhfah & \arb{۾}   \\
		sakt & \arb{ۜ}  \\
		dama\_mokhtalasa & \arb{ؙ}  \\
		\bottomrule
	\end{tabular}
\end{table*}

\begin{table*}[t]
	\centering
	\caption{Sifat Set (10 Attributes)}
	\label{tab:sifat_set}
	\begin{tabularx}{\linewidth}{@{}>{\raggedright}p{0.22\linewidth}>{\centering\arraybackslash}p{0.18\linewidth}>{\raggedright}p{0.27\linewidth}>{\centering\arraybackslash}p{0.23\linewidth}@{}}
		\toprule
		\textbf{Sifat (English)} & \textbf{Sifat (Arabic)}  & \textbf{Available Attributes (English)} & \textbf{Available Attributes (Arabic)} \\
		\midrule
		hams\_or\_jahr           & \arb{الهمس أو الجهر}     & hams, jahr                              & \arb{همس, جهر}                         \\
		shidda\_or\_rakhawa      & \arb{الشدة أو الرخاوة}   & shadeed, between, rikhw                 & \arb{شديد, بين بين, رخو}               \\
		tafkheem\_or\_taqeeq     & \arb{التفخيم أو الترقيق} & mofakham, moraqaq, low\_mofakham                       & \arb{مفخم, مرقق, أدنى المفخم}                       \\
		itbaq                    & \arb{الإطباق}            & monfateh, motbaq                        & \arb{منفتح, مطبق}                      \\
		safeer                   & \arb{الصفير}             & safeer, no\_safeer                      & \arb{صفير, لا صفير}                    \\
		qalqla                   & \arb{القلقلة}            & moqalqal, not\_moqalqal                 & \arb{مقلقل, غير مقلقل}                 \\
		tikraar                  & \arb{التكرار}            & mokarar, not\_mokarar                   & \arb{مكرر, غير مكرر}                   \\
		tafashie                 & \arb{التفشي}             & motafashie, not\_motafashie             & \arb{متفشي, غير متفشي}                 \\
		istitala                 & \arb{الاستطالة}          & mostateel, not\_mostateel               & \arb{مستطيل, غير مستطيل}               \\
		ghonna                   & \arb{الغنة}              & maghnoon, not\_maghnoon                 & \arb{مغنون, غير مغنون}                 \\
		\bottomrule
	\end{tabularx}
\end{table*}

\subsection{Uthmani to Phonetic Conversion Operations}
\label{subsec:conversion_ops}
The 26 sequential phonetization operations:

\begin{enumerate}
    \item \textbf{DisassembleHrofMoqatta} (\arb{تفكيك حروف مقطعة}): Separates Quranic initials (e.g., \arb{الم، الر}) into individual letters.
    
    \item \textbf{SpecialCases} (\arb{حالات خاصة}): Handles special words like \arb{يبسط} that have different pronunciation forms defined in \texttt{MoshafAttributes}.
    
    \item \textbf{BeginWithHamzatWasl} (\arb{البدء بهمزة الوصل}): Processes words starting with connecting hamza (\arb{ٱ}) and converts it to hamza (\arb{ء}) with appropriate harakah for nouns and verbs.
    
    \item \textbf{BeginWithSaken} (\arb{البدء بساكن}): Manages words beginning with a consonant (sakin) like \arb{لْيَقْطَعْ}, as Arabic doesn't start utterances with consonants.
    
    \item \textbf{ConvertAlifMaksora} (\arb{تحويل الألف المقصورة}): Converts \arb{ى} in Uthmani script to either yaa (\arb{ي}) or alif (\arb{ا}) based on context.
    
    \item \textbf{NormalizeHmazat} (\arb{توحيد الهمزات}): Standardizes hamza forms (\arb{أ إ ؤ ئ}) to \arb{ء}.
    
    \item \textbf{IthbatYaaYohie} (\arb{إثبات ياء يحيى}): Handles words like \arb{يُحْىِۦ} where two yaa letters occur - resolves conflicts when pausing on words with consecutive consonants (\arb{التقاء الساكنين}) by adding another yaa at end.
    
    \item \textbf{RemoveKasheeda} (\arb{إزالة الكشيدة}): Deletes elongation marks (\arb{ـــ}) from text.
    
    \item \textbf{RemoveHmzatWaslMiddle} (\arb{إزالة همزة الوصل الوسطية}): Removes connecting hamza (\arb{ٱ}) in non-initial positions.
    
    \item \textbf{RemoveSkoonMostadeer} (\arb{حذف الحرف الذي فوقع سكون مستدير}): Eliminates letters with circular sukoon diacritics like alif in \arb{جَمَعُوا۟}.
    
    \item \textbf{SkoonMostateel} (\arb{سكون مستطيل}): Removes alif with elongated sukoon mid-word and adds it at the end during pauses (\arb{وقف}).
    
    \item \textbf{MaddAlewad} (\arb{مد العوض}): Removes alif after tanween fatha mid-word and adds alif while removing tanween at pause positions (\arb{وقف}).
    
    \item \textbf{WawAlsalah} (\arb{واو الصلاة}): Replaces letter waw (\arb{و}) with small alif above combined with alif.
    
    \item \textbf{EnlargeSmallLetters} (\arb{تكبير الحروف الصغيرة}): Resizes miniature Arabic letters to standard proportions.
    
    \item \textbf{CleanEnd} (\arb{تنظيف النهاية}): Removes redundant diacritics and spaces at word endings.
    
    \item \textbf{NormalizeTaa} (\arb{توحيد التاء}): Converts \arb{ة} (taa marbuta) to \arb{ت} or \arb{ه} based on context, and converts final \arb{ة} to haa (\arb{ه}).
    
    \item \textbf{AddAlifIsmAllah} (\arb{إضافة ألف اسم الله}): Inserts compensatory alif in derivatives of "\arb{الله}".
    
    \item \textbf{PrepareGhonnaIdghamIqlab} (\arb{تهيئة الغنة والإدغام والإقلاب}): Preprocesses text for nasalization, assimilation, and conversion rules.
    
    \item \textbf{IltiqaaAlsaknan} (\arb{التقاء الساكنين}): Resolves consecutive consonants by inserting vowels.
    
    \item \textbf{DeleteShaddaAtBeginning} (\arb{حذف الشدة في البداية}): Removes shadda (\arb{ّ}) from word-initial letters.
    
    \item \textbf{Ghonna} (\arb{غنة}): Applies nasalization during pronunciation of sakin noon and tanween.
    
    \item \textbf{Tasheel} (\arb{تسهيل}): Adds a letter representing alif with tasheel easing.
    
    \item \textbf{Imala} (\arb{إمالة}): Converts fatha with imala to \texttt{fatha\_momala} phoneme and alif with imala to \texttt{alif\_momala} phoneme.
    
    \item \textbf{Madd} (\arb{مد}): Adds madd symbols for all madd types, inserting \texttt{madd\_alif} (\arb{ا}), \texttt{madd\_waw} (\arb{ۥ}), and \texttt{madd\_yaa} (\arb{ۦ}).
    
    \item \textbf{Qalqla} (\arb{قلقة}): Adds echoing effect to \arb{ق, ط, ب, ج, د} letters with sukoon.
    
    \item \textbf{RemoveRasHaaAndShadda} (\arb{إزالة رأس الحاء علامة السكون}): Deletes sukoon diacritic marks.
\end{enumerate}

\subsection{Tasmeea Verfication Algorithm}

\begin{figure*}[!htbp]
\caption{Tasmeea Algorithm}
\label{alg:tasmeea}
\begin{verbatim}
Algorithm 1 Tasmeea Algorithm
Input: text_segments, sura_idx, overlap_words=6,
       window_words=30, acceptance_ratio=0.85
Output: List of (match, ratio)
1: aya <- 1
2: penalty <- 0
3: for each segment s_i do
4:     norm_text <- normalize(s_i)
5:     best_ratio <- 0, best_match <- null
6:     for each start p in start_range do
7:         for each window w in [min_win, max_win] do
8:             candidate <- extract(aya, p, w)
9:             dist <- edit_distance(norm_text, candidate)
10:            ratio <- 1 - min(dist, |norm_text|)/|norm_text|
11:            if ratio > best_ratio then update best
12:         end for
13:     end for
14:     if best_ratio <= acceptance_ratio then
15:         output (null, best_ratio); penalty <- max_win; aya <- aya+1
16:     else
17:         output (best_match, best_ratio)
18:         aya <- aya + best_start + best_window; penalty <- 0
19:     end if
20: end for
\end{verbatim}
\end{figure*}

\subsection{Moshaf Attribute Definitions}
\label{sec:moshaf_attributes}
\begin{itemize}
\item \textbf{rewaya} (\arb{الرواية})
  \begin{itemize}
  \item Values: - \texttt{hafs} (\arb{حفص})
  \item Default Value: 
  \item More Info: The type of the quran Rewaya.
  \end{itemize}

\item \textbf{recitation\_speed} (\arb{سرعة التلاوة})
  \begin{itemize}
  \item Values: 
    \begin{itemize}
    \item  \texttt{mujawad} (\arb{مجود})
    \item  \texttt{above\_murattal} (\arb{فويق المرتل})
    \item  \texttt{murattal} (\arb{مرتل})
    \item  \texttt{hadr} (\arb{حدر})
    \end{itemize}
  \item Default Value: \texttt{murattal} (\arb{مرتل})
  \item More Info: The recitation speed sorted from slowest to the fastest \arb{سرعة التلاوة مرتبة من الأبطأ إلي الأسرع}
  \end{itemize}

\item \textbf{takbeer} (\arb{التكبير})
  \begin{itemize}
  \item Values: 
    \begin{itemize}
    \item  \texttt{no\_takbeer} (\arb{لا تكبير})
    \item  \texttt{beginning\_of\_sharh} (\arb{التكبير من أول الشرح لأول الناس})
    \item  \texttt{end\_of\_doha} (\arb{التكبير من آخر الضحى لآخر الناس})
    \item  \texttt{general\_takbeer} (\arb{التكبير أول كل سورة إلا التوبة})
    \end{itemize}
  \item Default Value: \texttt{no\_takbeer} (\arb{لا تكبير})
  \item More Info: The ways to add takbeer (\arb{الله أكبر}) after Istiaatha (\arb{استعاذة}) and between end of the surah and beginning of the surah. \texttt{no\_takbeer}: "\arb{لا تكبير}" — No Takbeer (No proclamation of greatness, i.e., there is no Takbeer recitation) \texttt{beginning\_of\_sharh}: "\arb{التكبير من أول الشرح لأول الناس}" — Takbeer from the beginning of Surah Ash-Sharh to the beginning of Surah An-Nas \texttt{end\_of\_dohaf}: "\arb{التكبير من آخر الضحى لآخر الناس}" — Takbeer from the end of Surah Ad-Duha to the end of Surah An-Nas \texttt{general\_takbeer}: "\arb{التكبير أول كل سورة إلا التوبة}" — Takbeer at the beginning of every Surah except Surah At-Tawbah
  \end{itemize}

\item \textbf{madd\_monfasel\_len} (\arb{مد المنفصل})
  \begin{itemize}
  \item Values: 
    \begin{itemize}
    \item  \texttt{2}
    \item  \texttt{3}
    \item  \texttt{4}
    \item  \texttt{5}
    \end{itemize}
  \item Default Value: 
  \item More Info: The length of Mad Al Monfasel "\arb{مد المنفصل}" for Hafs Rewaya.
  \end{itemize}

\item \textbf{madd\_mottasel\_len} (\arb{مقدار المد المتصل})
  \begin{itemize}
  \item Values: 
    \begin{itemize}
    \item  \texttt{4}
    \item  \texttt{5}
    \item  \texttt{6}
    \end{itemize}
  \item Default Value: 
  \item More Info: The length of Mad Al Motasel "\arb{مد المتصل}" for Hafs.
  \end{itemize}

\item \textbf{madd\_mottasel\_waqf} (\arb{مقدار المد المتصل وقفا})
  \begin{itemize}
  \item Values: 
    \begin{itemize}
    \item  \texttt{4}
    \item  \texttt{5}
    \item  \texttt{6}
    \end{itemize}
  \item Default Value: 
  \item More Info: The length of Madd Almotasel at pause for Hafs.. Example "\arb{السماء}".
  \end{itemize}

\item \textbf{madd\_aared\_len} (\arb{مقدار المد العارض})
  \begin{itemize}
  \item Values: 
    \begin{itemize}
    \item  \texttt{2}
    \item  \texttt{4}
    \item  \texttt{6}
    \end{itemize}
  \item Default Value: 
  \item More Info: The length of Mad Al Aared "\arb{مد العارض للسكون}".
  \end{itemize}

\item \textbf{madd\_alleen\_len} (\arb{مقدار مد اللين})
  \begin{itemize}
  \item Values: 
    \begin{itemize}
    \item  \texttt{2}
    \item  \texttt{4}
    \item  \texttt{6}
    \end{itemize}
  \item Default Value: \texttt{None}
  \item More Info: The length of the Madd al-Leen when stopping at the end of a word (for a sakin waw or ya preceded by a letter with a fatha) should be less than or equal to the length of Madd al-'Arid (the temporary stretch due to stopping). \textbf{Default Value is equal to \texttt{madd\_aared\_len}}. \arb{مقدار مد اللين عن القوف (للواو الساكنة والياء الساكنة وقبلها حرف مفتوح) ويجب أن يكون مقدار مد اللين أقل من أو يساوي مع العارض}
  \end{itemize}

\item \textbf{ghonna\_lam\_and\_raa} (\arb{غنة اللام و الراء})
  \begin{itemize}
  \item Values: 
    \begin{itemize}
    \item  \texttt{ghonna} (\arb{غنة})
    \item  \texttt{no\_ghonna} (\arb{لا غنة})
    \end{itemize}
  \item Default Value: \texttt{no\_ghonna} (\arb{لا غنة})
  \item More Info: The ghonna for merging (Idghaam) noon with Lam and Raa for Hafs.
  \end{itemize}

\item \textbf{meem\_aal\_imran} (\arb{ميم آل عمران في قوله تعالى: \{الم الله\} وصلا})
  \begin{itemize}
  \item Values: 
    \begin{itemize}
    \item  \texttt{waqf} (\arb{وقف})
    \item  \texttt{wasl\_2} (\arb{فتح الميم ومدها حركتين})
    \item  \texttt{wasl\_6} (\arb{فتح الميم ومدها ستة حركات})
    \end{itemize}
  \item Default Value: \texttt{waqf} (\arb{وقف})
  \item More Info: The ways to recite the word meem Aal Imran (\arb{الم الله}) at connected recitation. \texttt{waqf}: Pause with a prolonged madd (elongation) of 6 harakat (beats). \texttt{wasl\_2} Pronounce "meem" with fathah (a short "a" sound) and stretch it for 2 harakat. \texttt{wasl\_6} Pronounce "meem" with fathah and stretch it for 6 harakat.
  \end{itemize}

\item \textbf{madd\_yaa\_alayn\_alharfy} (\arb{مقدار   المد اللازم الحرفي للعين})
  \begin{itemize}
  \item Values: 
    \begin{itemize}
    \item  \texttt{2}
    \item  \texttt{4}
    \item  \texttt{6}
    \end{itemize}
  \item Default Value: \texttt{6}
  \item More Info: The length of Lzem Harfy of Yaa in letter Al-Ayen Madd "\arb{المد الحرفي اللازم لحرف العين}" in surar: Maryam "\arb{مريم}", AlShura "\arb{الشورى}".
  \end{itemize}

\item \textbf{saken\_before\_hamz} (\arb{الساكن قبل الهمز})
  \begin{itemize}
  \item Values: 
    \begin{itemize}
    \item  \texttt{tahqeek} (\arb{تحقيق})
    \item  \texttt{general\_sakt} (\arb{سكت عام})
    \item  \texttt{local\_sakt} (\arb{سكت خاص})
    \end{itemize}
  \item Default Value: \texttt{tahqeek} (\arb{تحقيق})
  \item More Info: The ways of Hafs for saken before hamz. "The letter with sukoon before the hamzah (\arb{ء})".And it has three forms: full articulation (\texttt{tahqeeq}), general pause (\texttt{general\_sakt}), and specific pause (\texttt{local\_skat}).
  \end{itemize}

\item \textbf{sakt\_iwaja} (\arb{السكت عند عوجا في الكهف})
  \begin{itemize}
  \item Values: 
    \begin{itemize}
    \item  \texttt{sakt} (\arb{سكت})
    \item  \texttt{waqf} (\arb{وقف})
    \item  \texttt{idraj} (\arb{إدراج})
    \end{itemize}
  \item Default Value: \texttt{waqf} (\arb{وقف})
  \item More Info: The ways to recite the word "\arb{عوجا}" (Iwaja). \texttt{sakt} means slight pause. \texttt{idraj} means not \texttt{sakt}. \texttt{waqf}: means full pause, so we can not determine whether the reciter uses \texttt{sakt} or \texttt{idraj} (no sakt).
  \end{itemize}

\item \textbf{sakt\_marqdena} (\arb{السكت عند مرقدنا  في يس})
  \begin{itemize}
  \item Values: 
    \begin{itemize}
    \item  \texttt{sakt} (\arb{سكت})
    \item  \texttt{waqf} (\arb{وقف})
    \item  \texttt{idraj} (\arb{إدراج})
    \end{itemize}
  \item Default Value: \texttt{waqf} (\arb{وقف})
  \item More Info: The ways to recite the word "\arb{مرقدنا}" (Marqadena) in Surat Yassen. \texttt{sakt} means slight pause. \texttt{idraj} means not \texttt{sakt}. \texttt{waqf}: means full pause, so we can not determine whether the reciter uses \texttt{sakt} or \texttt{idraj} (no sakt).
  \end{itemize}

\item \textbf{sakt\_man\_raq} (\arb{السكت عند  من راق في القيامة})
  \begin{itemize}
  \item Values: 
    \begin{itemize}
    \item  \texttt{sakt} (\arb{سكت})
    \item  \texttt{waqf} (\arb{وقف})
    \item  \texttt{idraj} (\arb{إدراج})
    \end{itemize}
  \item Default Value: \texttt{sakt} (\arb{سكت})
  \item More Info: The ways to recite the word "\arb{من راق}" (Man Raq) in Surat Al Qiyama. \texttt{sakt} means slight pause. \texttt{idraj} means not \texttt{sakt}. \texttt{waqf}: means full pause, so we can not determine whether the reciter uses \texttt{sakt} or \texttt{idraj} (no sakt).
  \end{itemize}

\item \textbf{sakt\_bal\_ran} (\arb{السكت عند  بل ران في  المطففين})
  \begin{itemize}
  \item Values: 
    \begin{itemize}
    \item  \texttt{sakt} (\arb{سكت})
    \item  \texttt{waqf} (\arb{وقف})
    \item  \texttt{idraj} (\arb{إدراج})
    \end{itemize}
  \item Default Value: \texttt{sakt} (\arb{سكت})
  \item More Info: The ways to recite the word "\arb{بل ران}" (Bal Ran) in Surat Al Motaffin. \texttt{sakt} means slight pause. \texttt{idraj} means not \texttt{sakt}. \texttt{waqf}: means full pause, so we can not determine whether the reciter uses \texttt{sakt} or \texttt{idraj} (no sakt).
  \end{itemize}

\item \textbf{sakt\_maleeyah} (\arb{وجه  قوله تعالى \{ماليه هلك\} بالحاقة})
  \begin{itemize}
  \item Values: 
    \begin{itemize}
    \item  \texttt{sakt} (\arb{سكت})
    \item  \texttt{waqf} (\arb{وقف})
    \item  \texttt{idgham} (\arb{إدغام})
    \end{itemize}
  \item Default Value: \texttt{waqf} (\arb{وقف})
  \item More Info: The ways to recite the word \{\arb{ماليه هلك}\} in Surah Al-Ahqaf. \texttt{sakt} means slight pause. \texttt{idgham} Assimilation of the letter 'Ha' (\arb{ه}) into the letter 'Ha' (\arb{ه}) with complete assimilation.\texttt{waqf}: means full pause, so we can not determine whether the reciter uses \texttt{sakt} or \texttt{idgham}.
  \end{itemize}

\item \textbf{between\_anfal\_and\_tawba} (\arb{وجه بين الأنفال والتوبة})
  \begin{itemize}
  \item Values: 
    \begin{itemize}
    \item  \texttt{waqf} (\arb{وقف})
    \item  \texttt{sakt} (\arb{سكت})
    \item  \texttt{wasl} (\arb{وصل})
    \end{itemize}
  \item Default Value: \texttt{waqf} (\arb{وقف})
  \item More Info: The ways to recite end of Surah Al-Anfal and beginning of Surah At-Tawbah.
  \end{itemize}

\item \textbf{noon\_and\_yaseen} (\arb{الإدغام والإظهار في النون عند الواو من قوله تعالى: \{يس والقرآن\}و \{ن والقلم\}})
  \begin{itemize}
  \item Values: 
    \begin{itemize}
    \item  \texttt{izhar} (\arb{إظهار})
    \item  \texttt{idgham} (\arb{إدغام})
    \end{itemize}
  \item Default Value: \texttt{izhar} (\arb{إظهار})
  \item More Info: Whether to merge noon of both: \{\arb{يس}\} and \{\arb{ن}\} with (\arb{و}) "\texttt{idgham}" or not "\texttt{izhar}".
  \end{itemize}

\item \textbf{yaa\_ataan} (\arb{إثبات الياء وحذفها وقفا في قوله تعالى \{آتان\} بالنمل})
  \begin{itemize}
  \item Values: 
    \begin{itemize}
    \item  \texttt{wasl} (\arb{وصل})
    \item  \texttt{hadhf} (\arb{حذف})
    \item  \texttt{ithbat} (\arb{إثبات})
    \end{itemize}
  \item Default Value: \texttt{wasl} (\arb{وصل})
  \item More Info: The affirmation and omission of the letter 'Yaa' in the pause of the verse \{\arb{آتاني}\} in Surah An-Naml. \texttt{wasl}: means connected recitation without pausing as (\arb{آتانيَ}). \texttt{hadhf}: means deletion of letter (\arb{ي}) at pause so recited as (\arb{آتان}). \texttt{ithbat}: means confirmation reciting letter (\arb{ي}) at pause as (\arb{آتاني}).
  \end{itemize}

\item \textbf{start\_with\_ism} (\arb{وجه البدأ بكلمة \{الاسم\} في سورة الحجرات})
  \begin{itemize}
  \item Values: 
    \begin{itemize}
    \item  \texttt{wasl} (\arb{وصل})
    \item  \texttt{lism} (\arb{لسم})
    \item  \texttt{alism} (\arb{ألسم})
    \end{itemize}
  \item Default Value: \texttt{wasl} (\arb{وصل})
  \item More Info: The ruling on starting with the word \{\arb{الاسم}\} in Surah Al-Hujurat. \texttt{lism} Recited as (\arb{لسم}) at the beginning. \texttt{alism} Recited as (\arb{ألسم}). \texttt{wasl}: means completing recitation without pausing as normal, So Reciting is as (\arb{بئس لسم}).
  \end{itemize}

\item \textbf{yabsut} (\arb{السين والصاد في قوله تعالى: \{والله يقبض ويبسط\} بالبقرة})
  \begin{itemize}
  \item Values: 
    \begin{itemize}
    \item  \texttt{seen} (\arb{سين})
    \item  \texttt{saad} (\arb{صاد})
    \end{itemize}
  \item Default Value: \texttt{seen} (\arb{سين})
  \item More Info: The ruling on pronouncing \texttt{seen} (\arb{س}) or \texttt{saad} (\arb{ص}) in the verse \{\arb{والله يقبض ويبسط}\} in Surah Al-Baqarah.
  \end{itemize}

\item \textbf{bastah} (\arb{السين والصاد في قوله تعالى:  \{وزادكم في الخلق بسطة\} بالأعراف})
  \begin{itemize}
  \item Values: 
    \begin{itemize}
    \item  \texttt{seen} (\arb{سين})
    \item  \texttt{saad} (\arb{صاد})
    \end{itemize}
  \item Default Value: \texttt{seen} (\arb{سين})
  \item More Info: The ruling on pronouncing \texttt{seen} (\arb{س}) or \texttt{saad} (\arb{ص}) in the verse \{\arb{وزادكم في الخلق بسطة}\} in Surah Al-A'raf.
  \end{itemize}

\item \textbf{almusaytirun} (\arb{السين والصاد في قوله تعالى \{أم هم المصيطرون\} بالطور})
  \begin{itemize}
  \item Values: 
    \begin{itemize}
    \item  \texttt{seen} (\arb{سين})
    \item  \texttt{saad} (\arb{صاد})
    \end{itemize}
  \item Default Value: \texttt{saad} (\arb{صاد})
  \item More Info: The pronunciation of \texttt{seen} (\arb{س}) or \texttt{saad} (\arb{ص}) in the verse \{\arb{أم هم المصيطرون}\} in Surah At-Tur.
  \end{itemize}

\item \textbf{bimusaytir} (\arb{السين والصاد في قوله تعالى:  \{لست عليهم بمصيطر\} بالغاشية})
  \begin{itemize}
  \item Values: 
    \begin{itemize}
    \item  \texttt{seen} (\arb{سين})
    \item  \texttt{saad} (\arb{صاد})
    \end{itemize}
  \item Default Value: \texttt{saad} (\arb{صاد})
  \item More Info: The pronunciation of \texttt{seen} (\arb{س}) or \texttt{saad} (\arb{ص}) in the verse \{\arb{لست عليهم بمصيطر}\} in Surah Al-Ghashiyah.
  \end{itemize}

\item \textbf{tasheel\_or\_madd} (\arb{همزة الوصل في قوله تعالى: \{آلذكرين\} بموضعي الأنعام و\{آلآن\} موضعي يونس و\{آلله\} بيونس والنمل})
  \begin{itemize}
  \item Values: 
    \begin{itemize}
    \item  \texttt{tasheel} (\arb{تسهيل})
    \item  \texttt{madd} (\arb{مد})
    \end{itemize}
  \item Default Value: \texttt{madd} (\arb{مد})
  \item More Info: Tasheel of Madd "\arb{وجع التسهيل أو المد}" for 6 words in The Holy Quran: "\arb{ءالذكرين}", "\arb{ءالله}", "\arb{ءائن}".
  \end{itemize}

\item \textbf{yalhath\_dhalik} (\arb{الإدغام وعدمه في قوله تعالى: \{يلهث ذلك\} بالأعراف})
  \begin{itemize}
  \item Values: 
    \begin{itemize}
    \item  \texttt{izhar} (\arb{إظهار})
    \item  \texttt{idgham} (\arb{إدغام})
    \item  \texttt{waqf} (\arb{وقف})
    \end{itemize}
  \item Default Value: \texttt{idgham} (\arb{إدغام})
  \item More Info: The assimilation (\texttt{idgham}) and non-assimilation (\texttt{izhar}) in the verse \{\arb{يلهث ذلك}\} in Surah Al-A'raf. \texttt{waqf}: means the reciter has paused on (\arb{يلهث})
  \end{itemize}

\item \textbf{irkab\_maana} (\arb{الإدغام والإظهار في قوله تعالى: \{اركب معنا\} بهود})
  \begin{itemize}
  \item Values: 
    \begin{itemize}
    \item  \texttt{izhar} (\arb{إظهار})
    \item  \texttt{idgham} (\arb{إدغام})
    \item  \texttt{waqf} (\arb{وقف})
    \end{itemize}
  \item Default Value: \texttt{idgham} (\arb{إدغام})
  \item More Info: The assimilation and clear pronunciation in the verse \{\arb{اركب معنا}\} in Surah Hud. This refers to the recitation rules concerning whether the letter "Noon" (\arb{ن}) is assimilated into the following letter or pronounced clearly when reciting this specific verse. \texttt{waqf}: means the reciter has paused on (\arb{اركب})
  \end{itemize}

\item \textbf{noon\_tamnna} (\arb{الإشمام والروم (الاختلاس) في قوله تعالى \{لا تأمنا على يوسف\}})
  \begin{itemize}
  \item Values: 
    \begin{itemize}
    \item  \texttt{ishmam} (\arb{إشمام})
    \item  \texttt{rawm} (\arb{روم})
    \end{itemize}
  \item Default Value: \texttt{ishmam} (\arb{إشمام})
  \item More Info: The nasalization (\texttt{ishmam}) or the slight drawing (\texttt{rawm}) in the verse \{\arb{لا تأمنا على يوسف}\}
  \end{itemize}

\item \textbf{harakat\_daaf} (\arb{حركة الضاد (فتح أو ضم) في قوله تعالى \{ضعف\} بالروم})
  \begin{itemize}
  \item Values: 
    \begin{itemize}
    \item  \texttt{fath} (\arb{فتح})
    \item  \texttt{dam} (\arb{ضم})
    \end{itemize}
  \item Default Value: \texttt{fath} (\arb{فتح})
  \item More Info: The vowel movement of the letter 'Dhad' (\arb{ض}) (whether with \texttt{fath} or \texttt{dam}) in the word \{\arb{ضعف}\} in Surah Ar-Rum.
  \end{itemize}

\item \textbf{alif\_salasila} (\arb{إثبات الألف وحذفها وقفا في قوله تعالى: \{سلاسلا\} بسورة الإنسان})
  \begin{itemize}
  \item Values: 
    \begin{itemize}
    \item  \texttt{hadhf} (\arb{حذف})
    \item  \texttt{ithbat} (\arb{إثبات})
    \item  \texttt{wasl} (\arb{وصل})
    \end{itemize}
  \item Default Value: \texttt{wasl} (\arb{وصل})
  \item More Info: Affirmation and omission of the 'Alif' when pausing in the verse \{\arb{سلاسلا}\} in Surah Al-Insan. This refers to the recitation rule regarding whether the final "Alif" in the word "\arb{سلاسلا}" is pronounced (affirmed) or omitted when pausing (waqf) at this word during recitation in the specific verse from Surah Al-Insan. \texttt{hadhf}: means to remove alif (\arb{ا}) during pause as (\arb{سلاسل}) \texttt{ithbat}: means to recite alif (\arb{ا}) during pause as (\arb{سلاسلا}) \texttt{wasl} means completing the recitation as normal without pausing, so recite it as (\arb{سلاسلَ وأغلالا})
  \end{itemize}

\item \textbf{idgham\_nakhluqkum} (\arb{إدغام القاف في الكاف إدغاما ناقصا أو كاملا \{نخلقكم\} بالمرسلات})
  \begin{itemize}
  \item Values: 
    \begin{itemize}
    \item  \texttt{idgham\_kamil} (\arb{إدغام كامل})
    \item  \texttt{idgham\_naqis} (\arb{إدغام ناقص})
    \end{itemize}
  \item Default Value: \texttt{idgham\_kamil} (\arb{إدغام كامل})
  \item More Info: Assimilation of the letter 'Qaf' into the letter 'Kaf,' whether incomplete (\texttt{idgham\_naqis}) or complete (\texttt{idgham\_kamil}), in the verse \{\arb{نخلقكم}\} in Surah Al-Mursalat.
  \end{itemize}

\item \textbf{raa\_firq} (\arb{التفخيم والترقيق في راء \{فرق\} في الشعراء وصلا})
  \begin{itemize}
  \item Values: 
    \begin{itemize}
    \item  \texttt{waqf} (\arb{وقف})
    \item  \texttt{tafkheem} (\arb{تفخيم})
    \item  \texttt{tarqeeq} (\arb{ترقيق})
    \end{itemize}
  \item Default Value: \texttt{tafkheem} (\arb{تفخيم})
  \item More Info: Emphasis and softening of the letter 'Ra' in the word \{\arb{فرق}\} in Surah Ash-Shu'ara' when connected (wasl). This refers to the recitation rules concerning whether the letter "Ra" (\arb{ر}) in the word "\arb{فرق}" is pronounced with emphasis (\texttt{tafkheem}) or softening (\texttt{tarqeeq}) when reciting the specific verse from Surah Ash-Shu'ara' in connected speech. \texttt{waqf}: means pausing so we only have one way (tafkheem of Raa)
  \end{itemize}

\item \textbf{raa\_alqitr} (\arb{التفخيم والترقيق في راء \{القطر\} في سبأ وقفا})
  \begin{itemize}
  \item Values: 
    \begin{itemize}
    \item  \texttt{wasl} (\arb{وصل})
    \item  \texttt{tafkheem} (\arb{تفخيم})
    \item  \texttt{tarqeeq} (\arb{ترقيق})
    \end{itemize}
  \item Default Value: \texttt{wasl} (\arb{وصل})
  \item More Info: Emphasis and softening of the letter 'Ra' in the word \{\arb{القطر}\} in Surah Saba' when pausing (waqf). This refers to the recitation rules regarding whether the letter "Ra" (\arb{ر}) in the word "\arb{القطر}" is pronounced with emphasis (\texttt{tafkheem}) or softening (\texttt{tarqeeq}) when pausing at this word in Surah Saba'. \texttt{wasl}: means not pausing so we only have one way (tarqeeq of Raa)
  \end{itemize}

\item \textbf{raa\_misr} (\arb{التفخيم والترقيق في راء \{مصر\} في يونس وموضعي يوسف والزخرف  وقفا})
  \begin{itemize}
  \item Values: 
    \begin{itemize}
    \item  \texttt{wasl} (\arb{وصل})
    \item  \texttt{tafkheem} (\arb{تفخيم})
    \item  \texttt{tarqeeq} (\arb{ترقيق})
    \end{itemize}
  \item Default Value: \texttt{wasl} (\arb{وصل})
  \item More Info: Emphasis and softening of the letter 'Ra' in the word \{\arb{مصر}\} in Surah Yunus, and in the locations of Surah Yusuf and Surah Az-Zukhruf when pausing (waqf). This refers to the recitation rules regarding whether the letter "Ra" (\arb{ر}) in the word "\arb{مصر}" is pronounced with emphasis (\texttt{tafkheem}) or softening (\texttt{tarqeeq}) at the specific pauses in these Surahs. \texttt{wasl}: means not pausing so we only have one way (tafkheem of Raa)
  \end{itemize}

\item \textbf{raa\_nudhur} (\arb{التفخيم والترقيق  في راء \{نذر\} بالقمر وقفا})
  \begin{itemize}
  \item Values: 
    \begin{itemize}
    \item  \texttt{wasl} (\arb{وصل})
    \item  \texttt{tafkheem} (\arb{تفخيم})
    \item  \texttt{tarqeeq} (\arb{ترقيق})
    \end{itemize}
  \item Default Value: \texttt{tafkheem} (\arb{تفخيم})
  \item More Info: Emphasis and softening of the letter 'Ra' in the word \{\arb{نذر}\} in Surah Al-Qamar when pausing (waqf). This refers to the recitation rules regarding whether the letter "Ra" (\arb{ر}) in the word "\arb{نذر}" is pronounced with emphasis (\texttt{tafkheem}) or softening (\texttt{tarqeeq}) when pausing at this word in Surah Al-Qamar. \texttt{wasl}: means not pausing so we only have one way (tarqeeq of Raa)
  \end{itemize}

\item \textbf{raa\_yasr} (\arb{التفخيم والترقيق في راء \{يسر\} بالفجر و\{أن أسر\} بطه والشعراء و\{فأسر\} بهود والحجر والدخان  وقفا})
  \begin{itemize}
  \item Values: 
    \begin{itemize}
    \item  \texttt{wasl} (\arb{وصل})
    \item  \texttt{tafkheem} (\arb{تفخيم})
    \item  \texttt{tarqeeq} (\arb{ترقيق})
    \end{itemize}
  \item Default Value: \texttt{tarqeeq} (\arb{ترقيق})
  \item More Info: Emphasis and softening of the letter 'Ra' in the word \{\arb{يسر}\} in Surah Al-Fajr when pausing (waqf). This refers to the recitation rules regarding whether the letter "Ra" (\arb{ر}) in the word "\arb{يسر}" is pronounced with emphasis (\texttt{tafkheem}) or softening (\texttt{tarqeeq}) when pausing at this word in Surah Al-Fajr. \texttt{wasl}: means not pausing so we only have one way (tarqeeq of Raa)
  \end{itemize}

\item \textbf{meem\_mokhfah} (\arb{هل الميم مخفاة أو مدغمة})
  \begin{itemize}
  \item Values: 
    \begin{itemize}
    \item  \texttt{meem} (\arb{ميم})
    \item  \texttt{ikhfaa} (\arb{إخفاء})
    \end{itemize}
  \item Default Value: \texttt{ikhfaa} (\arb{إخفاء})
  \item More Info: This is not a \textbf{standard} Hafs way but a disagreement between \textbf{scholars} in our century on how to \textbf{pronounce} \textbf{Ikhfa} for meem. Some \textbf{scholars} do full merging (\arb{إدغام}) and the others open the \textbf{lips} a little bit (\arb{إخفاء}). We did not want to add this, but some of the best reciters disagree about this.
  \end{itemize}
\end{itemize}

\subsection{qdat\_bench Dataset}
\label{sec:qdat_bench}

qdat\_bench is a comprehensive benchmark dataset for evaluating model performance in processing Quranic audio recordings with focus on Tajweed rules. This dataset builds upon the original qdat dataset \cite{osman2021qdat} after extensive reannotation and enhancement to include all Tajweed rules and QPS~\ref{sec:qps} (complete phoneme-level annotations and 10 sifat (characteristic) levels).

The enhanced dataset provides F1 and MSE metrics for Tajweed rules, enabling researchers to compare different representations and approaches to Tajweed rule detection. This comprehensive annotation framework makes qdat\_bench particularly valuable for advancing research in automated Quranic pronunciation assessment and error detection.

The dataset addresses several limitations of the original qdat collection. The original dataset suffered from incomplete coverage of all Tajweed rules, with only partial implementation of the comprehensive rule set required for thorough evaluation. Additionally, the original collection contained multiple reciters recording the same verse (each reciter records the same verse 10 times), creating significant redundancy and potential bias in evaluation. In contrast, qdat\_bench selects a single recording randomly for every reciter. Finally, the original dataset lacked comprehensive phoneme-level and sifat-level annotations, limiting its utility for fine-grained analysis of pronunciation patterns and characteristics.

qdat\_bench contains 159 samples focusing on the verse: \arb{قَالُوا۟ لَا عِلْمَ لَنَآ إِنَّكَ أَنتَ عَلَّٰمُ ٱلْغُيُوبِ} from Surah Al-Ma'idah (5:109), providing a concentrated evaluation of key Tajweed rules.

\subsubsection{Data Structure}
The dataset encompasses several key components designed for comprehensive analysis of Quranic recitation patterns. Each entry contains an audio file recorded in mono channel format, a unique identifiers for each element, reciter gender (male or female) and age, enabling analysis of pronunciation patterns across different population segments. The phonetic\_transcript and sifat fields contain the complete transcription in Quran Phonetic Script (QPS) as described in Section~\ref{sec:qps}, Along with Tajweed rules columns to enable different Tajweed representations benchmark on qdat\_bench. The dataset was annotated by a single hafiz of the Moshaf with Tajweed knowledge. Below is description of Tajweed rules found on our benchmark:

\paragraph{Madd (Prolongation) Rules:}
The dataset includes comprehensive annotations for various types of Madd rules, which are fundamental to proper Quranic recitation. The normal Madd rules are captured through several specific metrics: \texttt{qalo\_alif\_len} measures the length of normal Madd alif in the word \arb{قالوا} on a scale of 0-8, while \texttt{qalo\_waw\_len} similarly measures the normal Madd waw in the same word. Additional normal Madd measurements include \texttt{laa\_alif\_len} for the normal Madd alif in \arb{لا} and \texttt{allam\_alif\_len} for the normal Madd alif in \arb{علام}. The Separate Madd is measured through \texttt{separate\_madd}, which captures the length for the phrase \arb{لنا إنك} (0-8). Finally, Madd Aared, is evaluated using \texttt{madd\_aared\_len}, measuring prolongation before sukoon (0-8), which typically exhibits the highest variability in implementation.

\paragraph{Ghunnah (Nasalization) Rules:}
Ghunnah rules are systematically annotated to capture the nasalization characteristics essential for proper Quranic pronunciation. The \texttt{noon\_moshaddadah\_len} metric evaluates the length of noon moshaddadah in the word \arb{إنَّك} using a binary classification system where 0 indicates partial nasalization and 1 represents complete implementation. Similarly, the \texttt{noon\_mokhfah\_len} measures the Ikhfaa pronunciation in \arb{أنت} through a three-tiered system: 0 represents a clear noon pronunciation, 1 indicates partial Ikhfaa implementation, and 2 denotes complete Ikhfaa execution.

\paragraph{Qalqalah (Echo) Rules:}
The Qalqalah rule is captured through the \texttt{qalqalah} metric, which identifies the presence or absence of the echo characteristic in the word \arb{الغيوب}. This binary classification system uses 0 to indicate no Qalqalah implementation and 1 to denote proper Qalqalah execution.

\subsubsection{Dataset Statistics}

\begin{figure*}[!htbp]
	\centering
	\includegraphics[width=0.8\linewidth]{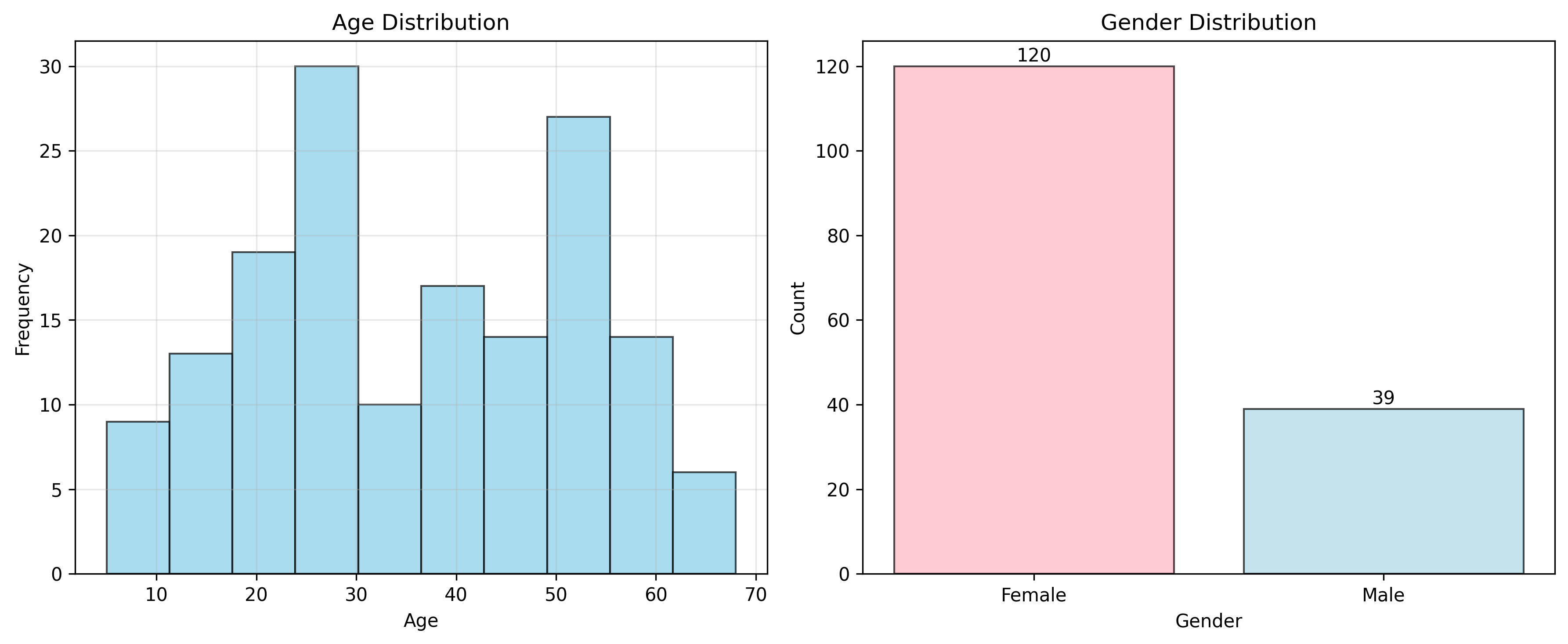}
	\caption{Age and gender distribution of qdat\_bench reciters, showing diverse demographic coverage with 75.5\% female and 24.5\% male participants across different age groups.}
	\label{fig:qdat_age_gender}
\end{figure*}

\begin{figure}[t]
	\centering
	\includegraphics[width=0.8\linewidth]{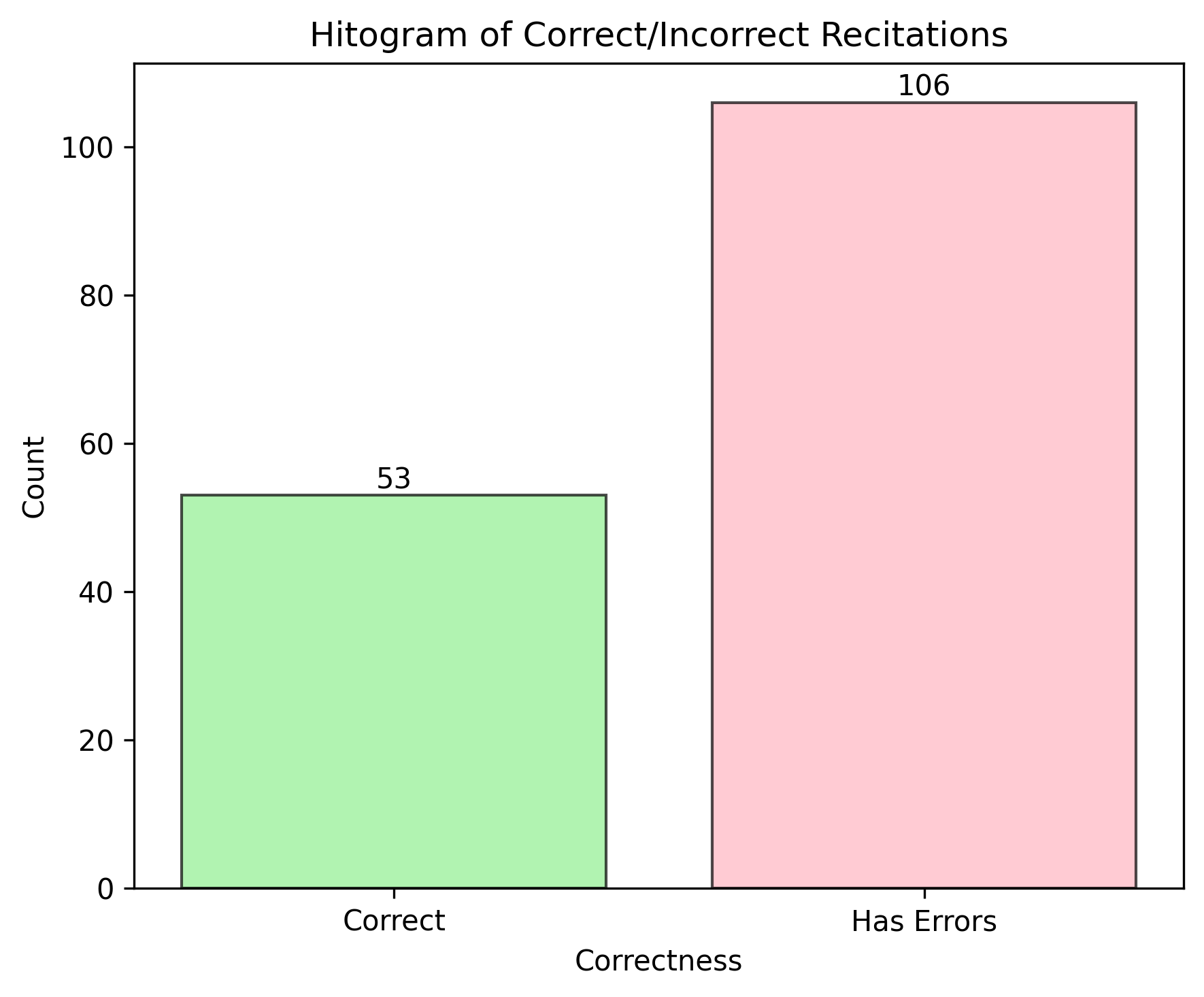}
	\caption{Distribution of recitation correctness across different Tajweed rules in the benchmark: red represents a reciter has one or more errors and green with correct recitation.}
	\label{fig:qdat_correctness}
\end{figure}

\begin{figure*}[!htbp]
	\centering
	\includegraphics[width=0.8\linewidth]{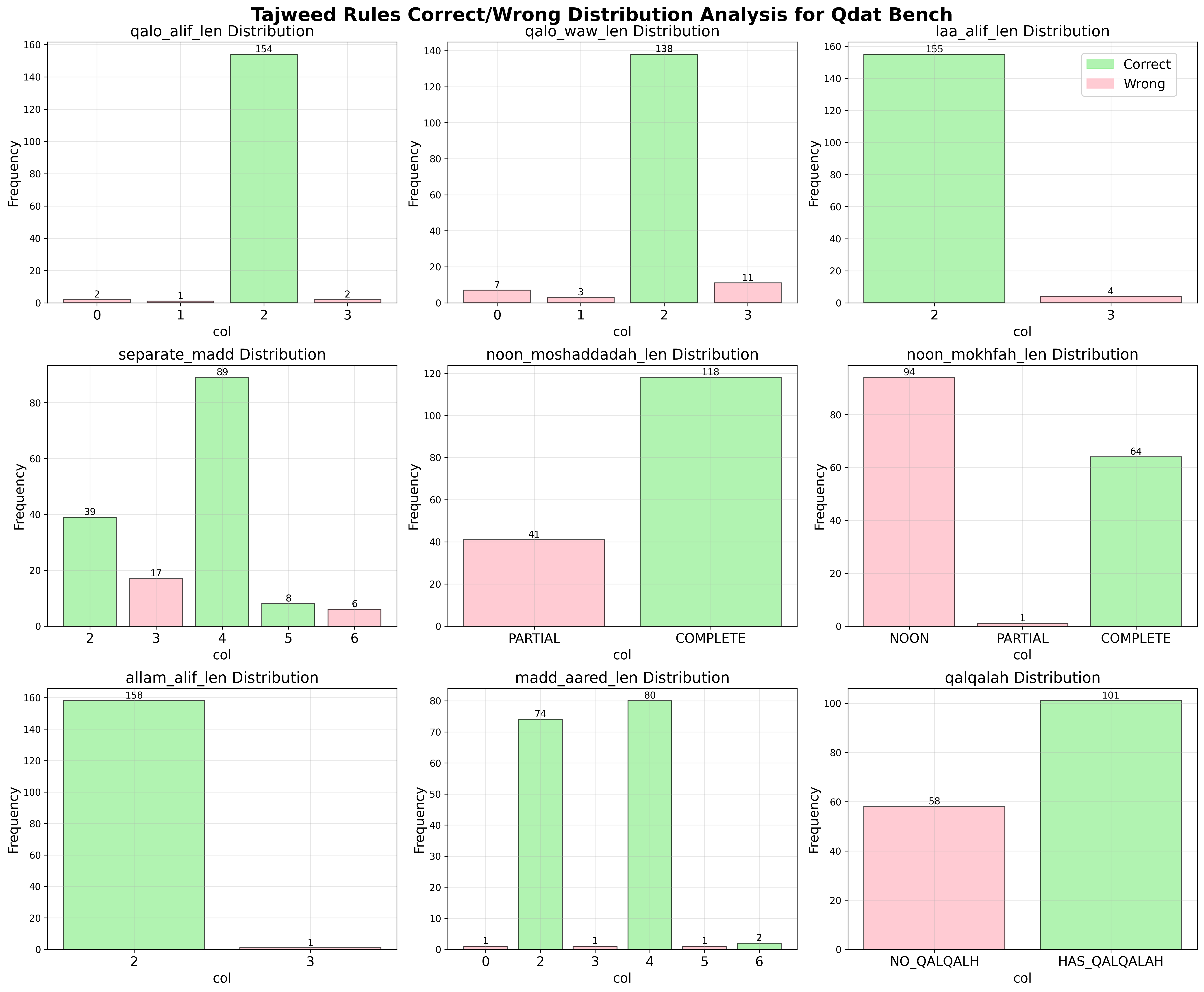}
	\caption{Tajweed rules coverage histogram showing the frequency and diversity of rules evaluated in the benchmark dataset, red bars represents errors and green with correct recitation.}
	\label{fig:qdat_tajweed_coverage}
\end{figure*}

The qdat\_bench dataset comprises 159 carefully selected samples. The demographic distribution reflects a diverse participant pool, with 120 female reciters representing 75.5\% of the dataset and 39 male reciters accounting for 24.5\%. The age diversity across various age groups provides comprehensive coverage of different learning stages and pronunciation patterns as shown in Figure~\ref{fig:qdat_age_gender}. The benchmark has various types of error: 106 reciters have one or more errors while 53 have complete correct recitations as shown in Figure~\ref{fig:qdat_correctness}. Finally Figure~\ref{fig:qdat_tajweed_coverage} shows the errors per every Tajweed rule as red represents errors and green with correct recitation.

\subsubsection{Evaluation Results}
The detailed evaluation results on qdat\_bench are presented in the following tables, showing performance across different Tajweed rule categories and metrics.

\begin{table*}[t]
	\centering
	\caption{Detailed qdat\_bench Speech Metrics. Note: Original refers to evaluation on the full test set without bootstrapping.}
	\label{tab:qdat_speech_metrics}
	\begin{tabular}{lcc}
		\hline
		\textbf{Metric}           & \textbf{Original} & \textbf{Bootstrap (n=10,000)} \\
		\hline
		per\_phonemes             & 0.0578            & 0.0603 $\pm$ 0.0053           \\
		per\_hams\_or\_jahr       & 0.0170            & 0.0172 $\pm$ 0.0033           \\
		per\_shidda\_or\_rakhawa  & 0.0310            & 0.0315 $\pm$ 0.0050           \\
		per\_tafkheem\_or\_taqeeq & 0.0224            & 0.0227 $\pm$ 0.0044           \\
		per\_itbaq                & 0.0117            & 0.0119 $\pm$ 0.0032           \\
		per\_safeer               & 0.0103            & 0.0103 $\pm$ 0.0028           \\
		per\_qalqla               & 0.0112            & 0.0112 $\pm$ 0.0026           \\
		per\_tikraar              & 0.0126            & 0.0128 $\pm$ 0.0027           \\
		per\_tafashie             & 0.0156            & 0.0159 $\pm$ 0.0035           \\
		per\_istitala             & 0.0093            & 0.0094 $\pm$ 0.0026           \\
		per\_ghonna               & 0.0140            & 0.0139 $\pm$ 0.0036           \\
		average\_per              & 0.0194            & 0.0197 $\pm$ 0.0032           \\
		\hline
	\end{tabular}
\end{table*}

\begin{table*}[t]
	\centering
	\caption{qdat\_bench Madd Rules Performance (RMSE). Golden values are for madd are: (2 for normal madd, 2 or 4 for separate madd, and 2, 4, or 6 for aared madd. Note: Original refers to evaluation on the full test set without bootstrapping.}
	\label{tab:qdat_madd_metrics}
	\begin{tabular}{lcc}
		\hline
		\textbf{Madd Rule}         & \textbf{Original} & \textbf{Bootstrap (n=10,000)} \\
		\hline
		qalo\_alif\_len (normal)   & 0.4486            & 0.4504 $\pm$ 0.0676           \\
		qalo\_waw\_len (normal)    & 0.4556            & 0.4609 $\pm$ 0.0561           \\
		laa\_alif\_len (normal)    & 0.4044            & 0.4065 $\pm$ 0.0713           \\
		separate\_madd             & 0.6868            & 0.6758 $\pm$ 0.0582           \\
		allam\_alif\_len (normal)  & 0.5494            & 0.5575 $\pm$ 0.0749           \\
		madd\_aared\_len           & 1.0340            & 1.0251 $\pm$ 0.0724           \\
		\hline
		\textbf{Average Madd RMSE} & 0.5965            & 0.5960 $\pm$ 0.0374           \\
		\hline
	\end{tabular}
\end{table*}

\begin{table*}[t]
	\centering
	\caption{qdat\_bench Noon Moshaddadah Performance. Note: Original refers to evaluation on the full test set without bootstrapping.}
	\label{tab:qdat_noon_moshaddadah}
	\begin{tabular}{lccc}
		\hline
		\textbf{Metric}               & \textbf{Partial}                        & \textbf{Complete}   & \textbf{Average}    \\
		\hline
		Recall                        & 0.6585                                  & 1.0000              & 0.8293              \\
		Precision                     & 1.0000                                  & 0.8939              & 0.9470              \\
		F1 Score                      & 0.7941                                  & 0.9440              & 0.8691              \\
		Accuracy                      & \multicolumn{3}{c}{0.9119}                                                          \\
		\hline
		\textbf{Bootstrap (n=10,000)} & \textbf{}                               & \textbf{}           & \textbf{}           \\
		Recall                        & 0.6604 $\pm$ 0.0721                     & 1.0000 $\pm$ 0.0000 & 0.8302 $\pm$ 0.0361 \\
		Precision                     & 1.0000 $\pm$ 0.0000                     & 0.8928 $\pm$ 0.0268 & 0.9464 $\pm$ 0.0134 \\
		F1 Score                      & 0.7932 $\pm$ 0.0531                     & 0.9431 $\pm$ 0.0150 & 0.8682 $\pm$ 0.0324 \\
		Accuracy                      & \multicolumn{3}{c}{0.9113 $\pm$ 0.0222}                                             \\
		\hline
	\end{tabular}
\end{table*}

\begin{table*}[t]
	\centering
	\caption{qdat\_bench Noon Mokhfah Performance. Note: Original refers to evaluation on the full test set without bootstrapping.}
	\label{tab:qdat_noon_mokhfah}
	\begin{tabular}{lccc}
		\hline
		\textbf{Metric}               & \textbf{Noon}                           & \textbf{Partial}    & \textbf{Complete}   \\
		\hline
		Recall                        & 0.4681                                  & 0.0000              & 0.9844              \\
		Precision                     & 1.0000                                  & 0.0000              & 0.5676              \\
		F1 Score                      & 0.6377                                  & 0.0000              & 0.7200              \\
		Average F1                    & \multicolumn{3}{c}{0.4526}                                                          \\
		Accuracy                      & \multicolumn{3}{c}{0.6730}                                                          \\
		\hline
		\textbf{Bootstrap (n=10,000)} & \textbf{}                               & \textbf{}           & \textbf{}           \\
		Recall                        & 0.4721 $\pm$ 0.0539                     & 0.0000 $\pm$ 0.0000 & 0.9820 $\pm$ 0.0165 \\
		Precision                     & 1.0000 $\pm$ 0.0000                     & 0.0000 $\pm$ 0.0000 & 0.5617 $\pm$ 0.0458 \\
		F1 Score                      & 0.6395 $\pm$ 0.0504                     & 0.0000 $\pm$ 0.0000 & 0.7134 $\pm$ 0.0371 \\
		Average F1                    & \multicolumn{3}{c}{0.4510 $\pm$ 0.0259}                                           \\
		Accuracy                      & \multicolumn{3}{c}{0.6716 $\pm$ 0.0376}                                           \\
		\hline
	\end{tabular}
\end{table*}

\begin{table*}[t]
	\centering
	\caption{qdat\_bench Qalqalah Performance. Note: Original refers to evaluation on the full test set without bootstrapping.}
	\label{tab:qdat_qalqalah}
	\begin{tabular}{lcc}
		\hline
		\textbf{Metric}               & \textbf{No Qalqalah}                    & \textbf{Has Qalqalah} \\
		\hline
		Recall                        & 0.9655                                  & 0.9505                \\
		Precision                     & 0.9180                                  & 0.9796                \\
		F1 Score                      & 0.9412                                  & 0.9648                \\
		Macro F1                      & \multicolumn{2}{c}{0.9530}                                      \\
		Accuracy                      & \multicolumn{2}{c}{0.9560}                                      \\
		\hline
		\textbf{Bootstrap (n=10,000)} & \textbf{}                               & \textbf{}             \\
		Recall                        & 0.9611 $\pm$ 0.0260                     & 0.9532 $\pm$ 0.0217   \\
		Precision                     & 0.9244 $\pm$ 0.0339                     & 0.9765 $\pm$ 0.0156   \\
		F1 Score                      & 0.9419 $\pm$ 0.0219                     & 0.9645 $\pm$ 0.0136   \\
		Macro F1                      & \multicolumn{2}{c}{0.9532 $\pm$ 0.0174}                         \\
		Accuracy                      & \multicolumn{2}{c}{0.9562 $\pm$ 0.0164}                         \\
		\hline
	\end{tabular}
\end{table*}

\subsubsection{Usage}
The dataset can be loaded using the Hugging Face datasets library:

\begin{verbatim}
from datasets import load_dataset
ds = load_dataset('obadx/qdat_bench')
print(ds['train'][0])  # Display first sample
\end{verbatim}

\subsection{Quran Phonetic Script Construction}

The Quran Phonetic Script is a set of letters and attributes (\arb{صفات}) that describes what the Holy Quran's reciters \textbf{actually} said. It was designed to capture all recitation rules, including all Tajweed rules (except Ishmam \arb{إشمام} and pausing with rawm \arb{روم} or \arb{إشمام}) and Sifat. This script is composed of 11 levels:

\begin{itemize}
	\item \texttt{phonemes} level: Designed to capture pronunciation of letters like baa (\arb{ب}) and diaractization like (fatha, damma and kasra).
	\item \texttt{sifat} level: Consisting of 10 levels to capture the attribute of articulation (\arb{صفة}) for every phoneme group.
\end{itemize}

We built this script based on \texttt{Hafs} (\arb{رواية حفص}) and incorporated all the different ways of reciting for \texttt{Hafs}. For example, the length of Madd Almunfasil can be (2, 3, 4, or 5 beats). Other variations can be found in the Hafs recitation methods.

\subsubsection{Phonemes Level}

The phoneme level has specific features, which are summarized as:

\begin{enumerate}
	\item \textbf{Madd Representation}:
	      \begin{itemize}
		      \item Normal Madd appears as consecutive madd symbols (e.g., 4-beat Madd: \arb{اااا}).
		      \item Madd al-Leen is represented with multiple waw/yaa symbols.
	      \end{itemize}

	\item \textbf{Ghunnah}:
	      \begin{itemize}
		      \item Stressed Ghunnah for noon (e.g., \arb{النون المشددة}) is represented as three consecutive noon symbols (\arb{ننن}).
		      \item Ikhfa is represented as three consecutive noon\_mokhfah (\arb{ںںں}) or meem\_mokhfah (\arb{۾۾۾}).
	      \end{itemize}

	\item \textbf{Idgham Handling}:
	      \begin{itemize}
		      \item Idgham for sakin noon with yaa is represented by consonant doubling (e.g., \arb{مَن يَعْمَلْ} → \arb{مَيييَعمَل}).
	      \end{itemize}

	\item \textbf{Special Cases}:
	      \begin{itemize}
		      \item Sakin: No following vowel symbol.
		      \item Imala: Represented by fatha\_momala and alif\_momala.
		      \item Rawm: Represented by the dama\_mokhtalasa marker.
	      \end{itemize}
\end{enumerate}

\begin{table*}[h]
	\centering
	\caption{Examples of Uthmani to Phonetic Script Conversion with Sifat Attributes}
	\label{tab:examples_with_sifat_appendix}
	\scriptsize
	\setlength{\tabcolsep}{3pt}
	\begin{tabular}{@{}>{\centering\arraybackslash}m{1.2cm} >{\centering\arraybackslash}m{1.5cm} *{10}{>{\centering\arraybackslash}m{0.7cm}@{}}}
		\toprule
		\textbf{Uthmani} & \textbf{Phonetic} &
		\textbf{H/J}     &
		\textbf{S/R}     &
		\textbf{T/T}     &
		\textbf{Itb}     &
		\textbf{Saf}     &
		\textbf{Qal}     &
		\textbf{Tik}     &
		\textbf{Taf}     &
		\textbf{Ist}     &
		\textbf{Gho}                                                                                     \\
		\cmidrule(lr){1-1} \cmidrule(lr){2-2} \cmidrule(lr){3-12}
		\arb{أَ}          & \arb{ءَ}           & jahr & shd & mrq & mnf & no & nql & nkr & ntf & nst & nmg \\
		\arb{تُ}          & \arb{تُ}           & hams & shd & mrq & mnf & no & nql & nkr & ntf & nst & nmg \\
		\arb{حَـ}         & \arb{حَ}           & hams & rkh & mrq & mnf & no & nql & nkr & ntf & nst & nmg \\
		\arb{ـٰٓ}          & \arb{اااااا}      & hams & rkh & mrq & mnf & no & nql & nkr & ntf & nst & nmg \\
		\arb{جُّ}          & \arb{ججُ}          & jahr & shd & mrq & mnf & no & nql & nkr & ntf & nst & nmg \\
		\arb{وٓ}          & \arb{ۥۥۥۥۥۥ}      & jahr & rkh & mrq & mnf & no & nql & nkr & ntf & nst & nmg \\
		\arb{نِّ}          & \arb{ننننِ}        & jahr & btw & mrq & mnf & no & nql & nkr & ntf & nst & mg  \\
		\arb{ى}          & \arb{ۦۦ}          & jahr & rkh & mrq & mnf & no & nql & nkr & ntf & nst & nmg \\
		\bottomrule
	\end{tabular}

	\vspace{2mm}
	\begin{center}  
		\scriptsize
		Phonetization of word (\arb{أَتُحَٰٓجُّوٓنِّى}) \\
		\textbf{Attribute Abbreviations:} \\
		H/J: Hams/Jahr \quad S/R: Shidda/Rakhawa \quad T/T: Tafkheem/Taqeeq \quad Itb: Itbaq \\
		Saf: Safeer \quad Qal: Qalqla \quad Tik: Tikraar \quad Taf: Tafashie \quad Ist: Istitala \quad Gho: Ghonna \\

		\textbf{Value Abbreviations:} \\
		shd: shadeed \quad rkh: rikhw \quad btw: between \quad mrq: moraqaq \\
		mof: mofakham \quad mnf: monfateh \quad mtb: motbaq \quad no: no\_safeer \\
		nql: not\_moqalqal \quad nkr: not\_mokarar \quad ntf: not\_motafashie \\
		nst: not\_mostateel \quad nmg: not\_maghnoon \quad mg: maghnoon
	\end{center}  
\end{table*}

We only care about pronounced phonemes of letters. If a letter is dropped or not pronounced, we will omit it. For example, we drop the Wasl Hamza (\arb{همزة الوصل}) when it appears in a context like: (\arb{بِسْمِ اللَّهِ}).

\subsubsection{Disconnected Letters}

Disconnected letters (\arb{الحروف المقطعة}) are letters that are pronounced as individual alphabets one by one. For example: (\arb{الٓمٓ}) is pronounced (\arb{أَلِفْ لَآم مِّيٓمْ}). There are 14 forms of these disconnected letters, so we must separate them according to their actual pronunciation.

\subsubsection{Madd (\arb{المد})}

There are three types of elongation (\arb{مد}):
\begin{itemize}
	\item \textbf{Madd Alif} (\arb{مد ألف}): Fatha followed by alif (\arb{ا})
	\item \textbf{Madd Waw} (\arb{مد بالواو}): Damma followed by waw (\arb{و})
	\item \textbf{Madd Yaa} (\arb{مد ياء}): Kasra followed by Yaa (\arb{ي})
\end{itemize}

These Madd types have different lengths relative to the natural Madd (\arb{المد الطبيعي}). We created special symbols to denote each Madd type:

\begin{itemize}
	\item \textbf{Madd Alif} is denoted by multiple alif symbols (\arb{ا})
	\item \textbf{Madd Waw} is denoted by multiple small\_waw symbols, designated as \texttt{waw\_madd} (\arb{ۥ})
	\item \textbf{Madd Yaa} is denoted by multiple small\_yaa symbols, designated as \texttt{yaa\_madd} (\arb{ۦ})
\end{itemize}

\paragraph{Normal Madd (\arb{المد الطبيعي})}

Normal Madd is the type of elongation pronounced at its standard length without excessive prolongation. We denote it by doubling the respective \texttt{madd} phonemes. The example below \ref{tab:ex_normal_madd} shows all three types of Madd in a single word.

\begin{table*}[ht]
	\centering
	\caption{The table demonstrates the three types of normal Madd: Madd Alif (\arb{اا}), Madd Yaa (\arb{ۦۦ}), and Madd Waw (\arb{ۥۥ}), each represented with two symbols to indicate a two-beat elongation.}
	\label{tab:ex_normal_madd}
	\begin{tabular}{|c|c|}
		\hline
		\textbf{Uthmani Script} & \textbf{Phonetic Script} \\ \hline
		\arb{نُوحِيهَا}            & \arb{نُۥۥحِۦۦهَاا}          \\ \hline
	\end{tabular}
\end{table*}

\paragraph{Madd Small Silah (\arb{مد الصلة الصغرى})}

Along with Normal Madd, Small Silah Madd (\arb{مد الصلة الصغرى}) follows the same representation rules. For example \ref{tab:ex_small_silah}:

\begin{table*}[ht]
	\centering
	\caption{The table shows Small Silah Madd along with noon mushaddad denoted as 3 repeated \texttt{noon} (\arb{ن}) with a special qalqala sign: (\arb{ڇ}) for letter jeem (\arb{ج}).}
	\label{tab:ex_small_silah}
	\begin{tabular}{|c|c|}
		\hline
		\textbf{Uthmani Script}    & \textbf{Phonetic Script}           \\ \hline
		\arb{إِنَّهُۥ عَلَىٰ رَجْعِهِۦ لَقَادِرٌ} & \arb{ءِننننَهُۥۥ عَلَاا رَجڇعِهِۦۦ لَقَاادِر} \\ \hline
	\end{tabular}
\end{table*}

\paragraph{Madd Al-'Iwad (\arb{مد العوض})}

In addition, Madd Al-'Iwad (\arb{مد العوض}) is represented as shown in \ref{tab:ex_alewad}:

\begin{table*}[ht]
	\centering
	\caption{The table shows Madd Al-'Iwad (\arb{مد العوض}) using the same notation as normal Madd (\arb{المد الطبيعي}) for Madd alif. This type of Madd occurs when a tanween fatha on a final letter is replaced by an alif madd during pause.}
	\label{tab:ex_alewad}
	\begin{tabular}{|c|c|}
		\hline
		\textbf{Uthmani Script} & \textbf{Phonetic Script} \\ \hline
		\arb{قَرِيبًا}             & \arb{قَرِۦۦبَاا}            \\ \hline
	\end{tabular}
\end{table*}

\subsubsection{Madd Al-Munfasil (\arb{مد المنفصل})}

For Hafs recitation, Madd Al-Munfasil can be elongated for 2, 3, 4, or 5 harakat, where a haraka here is represented as half of a normal Madd when followed by a hamza (\arb{ء}) not in the same word, as shown in the example \ref{tab:ex_monfasel}:

\begin{table*}[ht]
	\centering
	\caption{The example shows elongation for Madd Al-Munfasil with 4 alif madd phonemes, along with a repeated yaa representing yaa mushaddada (\arb{ياء مشددة}) with both a sakin yaa and a yaa with haraka (damma).}
	\label{tab:ex_monfasel}
	\begin{tabular}{|c|c|}
		\hline
		\textbf{Uthmani Script} & \textbf{Phonetic Script} \\ \hline
		\arb{يَٰٓأَيُّهَا}             & \arb{يَااااءَييُهَاا}        \\ \hline
	\end{tabular}
\end{table*}

\paragraph{Madd As-Silah Al-Kubra (\arb{مد الصلة الكبرى})}

The same rule is applied to Madd As-Silah Al-Kubra (\arb{مد الصلة الكبرى}). As shown in the example below: \ref{tab:ex_big_silah}

\begin{table*}[ht]
	\centering
	\caption{The example shows elongation for Madd As-Silah Al-Kubra with 4 madd waw phonemes (\arb{ۥۥۥۥ}).}
	\label{tab:ex_big_silah}
	\begin{tabular}{|c|c|}
		\hline
		\textbf{Uthmani Script} & \textbf{Phonetic Script} \\ \hline
		\arb{مَالَهُۥٓ أَخْلَدَهُۥ}      & \arb{مَاالَهُۥۥۥۥ ءَخلَدَه}    \\ \hline
	\end{tabular}
\end{table*}

\paragraph{Madd Al-Muttasil (\arb{المد المتصل})}

For Hafs recitation, Madd Al-Muttasil (\arb{مد المتصل}) can be elongated for 2, 3, 4, 5, or (6 at pause only) harakat, where a haraka is represented as half of a normal Madd when followed by a hamza (\arb{ء}) in the same word, as shown in \ref{tab:ex_mottasel}:

\begin{table*}[ht]
	\centering
	\caption{The example shows elongation for Madd Al-Muttasil (\arb{مد المتصل}) with 4 madd alif phonemes, along with Madd Al-'Iwad (\arb{مد العوض}) at the pause point.}
	\label{tab:ex_mottasel}
	\begin{tabular}{|c|c|}
		\hline
		\textbf{Uthmani Script} & \textbf{Phonetic Script} \\ \hline
		\arb{ٱلسَّمَآءِ مَآءً}        & \arb{ءَسسَمَااااءِ مَااااءَاا} \\ \hline
	\end{tabular}
\end{table*}

\paragraph{Madd Al-Lazim (\arb{المد اللازم})}

Madd Al-Lazim (\arb{المد اللازم}) is the type of Madd where a Madd letter is followed by a Sakin letter (\arb{حرف ساكن}) in the same word and is elongated for 6 harakat (\arb{6 حركات}), where a haraka is represented as half of a normal Madd.

\begin{table*}[ht]
	\centering
	\caption{The table shows an example of Madd Al-Lazim (\arb{المد اللازم}) with Madd alif elongated for 6 harakat, along with Madd Al-'Arid Li-S-Sukun (\arb{مد العرض للسكون}) represented with 4 harakat.}
	\label{tab:ex_lazem}
	\begin{tabular}{|c|c|}
		\hline
		\textbf{Uthmani Script} & \textbf{Phonetic Script} \\ \hline
		\arb{ٱلضَّآلِّينَ}           & \arb{ءَضضَااااااللِۦۦۦۦنَ}   \\ \hline
	\end{tabular}
\end{table*}

\paragraph{Madd Al-\arb{عارض} Li-S-\arb{سكون} (\arb{مد العارض للسكون})}

Madd Al-\arb{عارض} Li-S-\arb{سكون} is the madd that occurs when pausing after a normal madd with a sakin letter. This madd is elongated for 2, 4, or 6 harakat, where the haraka is represented as half of the normal madd length, as shown in \ref{tab:ex_lazem}:

\paragraph{Madd Al-\arb{لين} (\arb{مد اللين})}

Madd Al-\arb{لين} (\arb{مد اللين}) occurs when pausing after a \arb{ياء} (\arb{ي}) or \arb{واو} (\arb{و}) that is preceded by a fatha and followed by a sakin letter. This madd is elongated for 2, 4, or 6 harakat, where a haraka is represented as half of the normal madd length \ref{tab:ex_leen}. We do not create special phonemes for this rule as we did with other madd types because \arb{لين} represents an elongation of existing \arb{واو} (\arb{و}) or \arb{ياء} (\arb{ي}) phonemes rather than introducing new phonemes.

For a 4-haraka madd, we denote this with (number\_of\_harakat - 1) symbols. This approach accounts for cases of Madd Al-\arb{لين} in the middle of recitation (like \arb{وَٱلْمَيْسِرِ}) as well as at pause positions, maintaining consistency in the phonetic script. Table \ref{tab:ex_leen} shows an example of Madd Al-\arb{لين}:

\begin{table*}[ht]
	\centering
	\caption{The example shows two forms of madd: the first is normal madd followed by Madd Al-\arb{لين} with 4 harakat (each haraka being half of normal madd), denoted with 3 \arb{ياء} (\arb{ي}) symbols.}
	\label{tab:ex_leen}
	\begin{tabular}{|c|c|}
		\hline
		\textbf{Uthmani Script} & \textbf{Phonetic Script} \\
		\hline
		\arb{لِإِيلَٰفِ قُرَيْشٍ}        & \arb{لِءِۦۦلَاافِ قُرَيييش}    \\
		\hline
	\end{tabular}
\end{table*}

\subsubsection{Ghunnah (\arb{العنة})}

We consider tanween here as a haraka (fatha, damma, or kasra) followed by a sakin noon (\arb{نون ساكنة}), so we do not need to define separate rules for noon (\arb{ن}) and tanween.

\paragraph{Noon Mushaddadah (\arb{النون المشددة})}

We first attempted to measure the relative timing of a sakin noon alone (\arb{النون الساكنة المظهرة}) and compare it to an elongated noon (noon with shaddah - \arb{نون مشددة}). We found that the elongated noon is approximately 3 to 4 times longer than the sakin noon, so we defined the elongated noon as equivalent to 3 sakin noon repetitions. Example in table \ref{tab:ex_noon_moshadada}

\begin{table*}[ht]
	\centering
	\caption{The table shows how Ghunnah disassembly of noon with shaddah (\arb{نون مشددة}) is represented as 3 repetitive noon (\arb{ن}) symbols.}
	\label{tab:ex_noon_moshadada}
	\begin{tabular}{|c|c|}
		\hline
		\textbf{Uthmani Script} & \textbf{Phonetic Script} \\
		\hline
		\arb{إِنَّ}                & \arb{ءِننن}               \\
		\hline
		\arb{شَىْءٍ نُّكُرٍ}           & \arb{شَيءِننننُكُر}          \\
		\hline
	\end{tabular}
\end{table*}

\paragraph{Meem Mushaddadah (\arb{الميم المشددة})}

As we have done with Noon Mushaddadah, we applied the same principle to Meem Mushaddadah (elongated meem). We found the same result: Meem Mushaddadah is approximately 3 to 4 times longer than a regular sakin meem (\arb{ميم ساكنة مظهرة}). We denote Meem Mushaddadah as 3 repeated meem symbols, as shown in the examples: \ref{tab:ex_meem_moshadda}

\begin{table*}[ht]
	\centering
	\caption{The table shows how Ghunnah disassembly of meem with shaddah (\arb{ميم مشددة}) is represented as 3 repeated meem (\arb{م}) symbols.}
	\label{tab:ex_meem_moshadda}
	\begin{tabular}{|c|c|}
		\hline
		\textbf{Uthmani Script} & \textbf{Phonetic Script} \\
		\hline
		\arb{أَمَّا}               & \arb{ءَممممَاا}            \\
		\hline
		\arb{خَيْرٍ مِّن}            & \arb{خَيرِممممِن}           \\
		\hline
	\end{tabular}
\end{table*}

\paragraph{Ikhfaa for Noon (\arb{إخفاء النون الساكنة})}

Ikhfaa for sakin noon (\arb{إخفاء النون الساكنة}) occurs when a sakin noon (\arb{نون ساكنة}) or tanween is followed by any of the Ikhfaa letters: (\arb{ص، ذ، ث، ك، ج، ش، ق، س، د، ط، ز، ت، ض، ظ، ف}). We denote this by replacing the noon with three \texttt{noon\_mokhfaa} symbols (\arb{ں}), as shown in the example \ref{tab:ex_noon_mokhfaa}:

\begin{table*}[ht]
	\centering
	\caption{The table shows the representation of noon mokhfaa (\arb{نون مخفاة}) as three dotless noon symbols (\arb{ں}).}
	\label{tab:ex_noon_mokhfaa}
	\begin{tabular}{|c|c|}
		\hline
		\textbf{Uthmani Script} & \textbf{Phonetic Script} \\
		\hline
		\arb{مِن صَلْصَٰلٍ}           & \arb{مِںںںصَلصَاااال}       \\
		\hline
	\end{tabular}
\end{table*}

\paragraph{Idgham for Noon with Yaa and Waw (\arb{إدغام النون الساكنة مع الياء والواو})}

The Idgham rule is defined as pronouncing two consecutive letters as the second letter with shadda (stress) according to Ibn Al-Jazari \cite{ibnaljazri_alnashr}. Therefore, we simply delete the noon (\arb{ن}) and replace it with a yaa (\arb{ي}) or waw (\arb{و}).

As with Noon Mushaddadah and Meem Mushaddadah, we represent the resulting stressed yaa or waw with two repetitions rather than three. This maintains consistency with our representation of Madd Al-\arb{لين} and follows the convention that a stressed letter (\arb{مشدد}) is represented by both a sakin and mutaharrik (\arb{متحرك}) form. Table \ref{tab:ex_idghaam_yaa_with_noon} shows examples of different forms of yaa:

\begin{table*}[ht]
	\centering
	\caption{This table demonstrates different representations of yaa. The first row shows Idgham of yaa with sakin noon (\arb{النون الساكنة}) represented by replacing the noon with two yaa symbols. The second row shows yaa with shadda at pause represented with two yaa symbols. The third row shows Madd Al-\arb{لين} with 4 harakat represented by 3 yaa symbols.}
	\label{tab:ex_idghaam_yaa_with_noon}
	\begin{tabular}{|c|c|}
		\hline
		\textbf{Uthmani Script} & \textbf{Phonetic Script} \\
		\hline
		\arb{مَن يَعْمَلْ}           & \arb{مَيييَعمَل}            \\
		\hline
		\arb{ٱلْحَىِّ}              & \arb{ءَلحَيي}              \\
		\hline
		\arb{قُرَيْشٍ}              & \arb{قُرَيييش}             \\
		\hline
	\end{tabular}
\end{table*}

\paragraph{Ikhfaa for Meem (\arb{إخفاء الميم الساكنة})}

Ikhfaa for sakin meem (\arb{إخفاء الميم الساكنة}) occurs when a sakin meem (\arb{ميم ساكنة}) is followed by a baa (\arb{ب}). Additionally, when a sakin noon or tanween is followed by baa, it is defined in Tajweed literature as Iqlab (\arb{إقلاب}). We represent both cases with three \texttt{meem\_mokhfah} symbols (\arb{۾}). Table \ref{tab:ex_ikhfaa_meem} shows how this rule is applied:

\begin{table*}[ht]
	\centering
	\caption{The first row represents the Iqlab rule (\arb{الإقلاب}), which is denoted by replacing the noon with 3 \texttt{meem\_mokhfah} symbols (\arb{۾}) and \arb{(ڇ)} donates Qalqala. The second row shows the rule of Ikhfaa for sakin meem with baa (\arb{إخفاء الميم الساكنة}), represented by 3 \texttt{meem\_mokhfah} symbols (\arb{۾}).}
	\label{tab:ex_ikhfaa_meem}
	\begin{tabular}{|c|c|}
		\hline
		\textbf{Uthmani Script} & \textbf{Phonetic Script} \\
		\hline
		\arb{مِنۢ بَعْدِ}            & \arb{مِ۾۾۾بَعدڇ}           \\
		\hline
		\arb{تَرْمِيهِم بِحِجَارَةٍ}     & \arb{تَرمِۦۦهِ۾۾۾بِحِجَاارَه}   \\
		\hline
	\end{tabular}
\end{table*}

\subsubsection{Idgham (\arb{الإدغام})}

There are two types of merging (Idgham) in Arabic:

\begin{itemize}
	\item \textbf{Full Merging (\arb{إدغام كامل})}: When two letters follow each other and are pronounced as only the second letter, but stressed. Example: (\arb{قَد تَّبَيَّنَ}) is pronounced as (\arb{قَتتَبَييَن}) where the letter daal is completely not pronounced.
	\item \textbf{Partial Merging (\arb{إدغام ناقص})}: When two letters follow each other and the articulation point (makhraj) of the first letter is lost but its attributes (sifat) remain. Example: (\arb{بَسَطْتَ}) is pronounced the same (\arb{بَسَطَت}).
\end{itemize}

\subsubsection{Sakin Letter (\arb{الحرف الساكن})}

A sakin letter is represented in the Uthmani script in three forms:

\begin{enumerate}
	\item A letter followed by sukun (\arb{سُكُون}): '\arb{ْ}'
	\item A letter with shaddah (\arb{شَدَّة}), which represents a stressed letter composed of two identical letters: the first is sakin and the second has a haraka (fattha, dammah, or kasrah). Example: (\arb{بَّ}) → (\arb{ببَ})
	\item A letter that is not followed by any haraka (short vowel) or any special symbol, which occurs in Idgham and with madd letters.
\end{enumerate}

We denote a sakin letter by the absence of any following vowel diacritic.

\subsubsection{Pausing (\arb{وَقْف})}

At a pause (\arb{وَقْف}), we make the final letter sakin (\arb{سَاكِن}) by removing any vowel diacritic. See examples in: \ref{tab:ex_idghaam_yaa_with_noon} and other relevant tables.

\subsubsection{Hamzat Al-Wasl (\arb{همزة الوصل})}

Hamzat Al-Wasl (\arb{همزة الوصل}) (\arb{ٱ}) is defined in Tajweed as a hamza added to avoid beginning with a sakin letter \cite{sweed2021}. It is elided during continuous recitation and is only pronounced at the beginning.

The vowel following Hamzat Al-Wasl (fatha, damma, or kasra) depends on the word type:

\begin{itemize}
	\item For nouns beginning with Alif-Lam at-ta'reef (\arb{ال التعريف}), the hamza is followed by fatha.
	\item For proper nouns, the hamza is followed by kasra.
	\item For verbs: the vowel depends on the third root letter:
	      \begin{itemize}
		      \item Damma: hamza is followed by non-transient damma
		      \item Fatha, kasra, or transient damma: hamza is followed by kasra
	      \end{itemize}
\end{itemize}

\textbf{Transient damma} refers to a damma that is not original but results from a temporary grammatical state. For example, the word (\arb{ٱمْشُوا۟}) has a damma on its third letter, but the verb originates from (\arb{ٱمْشِ}) where the third letter (\arb{ش}) has kasra.

\begin{table*}[ht]
	\centering
	\caption{This table shows different forms of Hamzat Al-Wasl (\arb{ٱ}). The first and second rows demonstrate beginning with hamza followed by fatha due to (\arb{ال}) at-ta'reef. The third row shows beginning with hamza followed by kasra for a proper noun. The 4th, 5th, and 6th rows show verbs beginning with hamza followed by kasra because the third radical has fatha, kasra, or transient damma. The last row shows beginning with hamza followed by damma because the third radical has a non-transient damma.}
	\label{tab:ex_hamzat_wasl}
	\begin{tabular}{|c|c|c|c|}
		\hline
		\textbf{Uthmani Script} & \textbf{Phonetic Script} & \textbf{Word Type}                        & \textbf{Hamzat Wasl Vowel} \\
		\hline
		\arb{ٱلْكِتَٰبُ}             & \arb{ءَلكِتَاااابڇ}         & Noun beginning with (\arb{ال})            & fatha                      \\
		\hline
		\arb{ٱللَّهِ}              & \arb{ءَللَااااه}           & Proper Noun beginning with (\arb{ال})     & fatha                      \\
		\hline
		\arb{ٱسْتِكْبَارًۭا}          & \arb{ءِستِكبَاارَاا}         & Proper Noun                               & kasra                      \\
		\hline
		\arb{ٱرْكَب}              & \arb{ءِركَبڇ}              & Verb (3rd letter has fatha)               & kasra                      \\
		\hline
		\arb{ٱصْبِرْ}              & \arb{ءِصبِر}               & Verb (3rd letter has kasra)               & kasra                      \\
		\hline
		\arb{ٱمْشُوا۟}             & \arb{ءِمشُۥۥ}              & Verb (3rd letter has transient damma)     & kasra                      \\
		\hline
		\arb{ٱرْكُضْ}              & \arb{ءُركُض}               & Verb (3rd letter has non-transient damma) & damma                      \\
		\hline
	\end{tabular}
\end{table*}

\textbf{Important Note}: We rely on Dukes's work \cite{dukes2010morphological} for determining word types (nouns, verbs, and particles). Without this foundational research, annotating the Holy Quran's words would require at least a year of dedicated effort, highlighting the critical importance of open-source linguistic resources.

\paragraph{Meeting Two Hamzas (Second One is Sakin) (\arb{التقاء همزتان والثانية منهما ساكنة})}

After converting Hamzat Wasl to a pronounced hamza, certain cases occur where two hamzas meet and the second one is sakin (consonant). In such cases, the second hamza is converted to a madd letter matching the vowel (haraka) of the first hamza \cite{sweed2021}. Table \ref{tab:ex_meeting_two_hamza} illustrates this process:

\begin{table*}[ht]
	\centering
	\caption{The table shows the conversion process for verbs that begin with two connected hamzas. The first stage converts Hamzat Wasl to a hamza followed by kasra or damma. The second stage converts the second hamza to either waw\_madd (\arb{ۥ}) or yaa\_madd (\arb{ۦ}), depending on the vowel of the first hamza. We maintain our established representation where normal madd is represented by two symbols: (\arb{اا}) for madd\_alif, (\arb{ۦۦ}) for madd\_yaa, and (\arb{ۥۥ}) for madd\_waw.}
	\label{tab:ex_meeting_two_hamza}
	\begin{tabular}{|c|c|c|}
		\hline
		\textbf{Uthmani Script} & \textbf{Converting Hamzat Wasl} & \textbf{Final Conversion} \\
		\hline
		\arb{ٱؤْتُمِنَ}             & \arb{ءُءْتُمِن}                     & \arb{ءُۥۥتُمِن}              \\
		\hline
		\arb{ٱئْتُونِى}            & \arb{ءِءْتُۥۥنِۦۦ}                  & \arb{ءِۦۦتُۥۥنِۦۦ}           \\
		\hline
	\end{tabular}
\end{table*}

\subsubsection{Meeting Two Sakin Letters (\arb{التقاء الساكنين})}

In Arabic language and the Holy Quran, two sakin letters (\arb{الحرفان الساكنان}) cannot meet consecutively except at pause (\arb{وقف}), such as pausing on the word (\arb{ٱلْأَرْضِ}) where the final two letters are sakin. To resolve this meeting, three approaches may be employed:

\begin{itemize}
	\item Eliminate the first letter
	\item Elongate the first letter
	\item Diacritize the second letter with a vowel (fatha, damma, or kasra)
\end{itemize}

Muslim scholars have simplified this task by comprehensively annotating these rules within the Uthmani script, except for two specific cases:
\begin{itemize}
	\item When the first letter is (alif, waw, or yaa): we eliminate the first letter
	\item When the first letter is tanween: we convert the tanween to a noon (\arb{ن}) followed by kasra
\end{itemize}

Table \ref{tab:ex_two_saken} shows how we apply this rule in our phonetization process:

\begin{table*}[ht]
	\centering
	\caption{The table demonstrates how we resolve the meeting of two sakin letters. The first row shows the meeting of alif (\arb{ا}) from the word (\arb{قَالَا}) with the lam (\arb{ل}) of the word (\arb{ٱلْحَمْدُ}). In the resulting phonetic script, the alif was deleted. Note that normal madd in (\arb{قَالَ}) is represented by two alif (\arb{اا}), and qalqala in the letter daal (\arb{د}) is represented by (\arb{ڇ}). The second example shows the meeting of tanween from (\arb{نُوحٌ}) with the sakin baa (\arb{ب}) of the word (\arb{ٱبْنَهُۥ}), resulting in the conversion of tanween to noon with kasra. Note also that normal madd waw is represented with two (\arb{ۥۥ}) and qalqala for baa (\arb{ب}) with (\arb{ڇ}).}
	\label{tab:ex_two_saken}
	\begin{tabular}{|c|c|}
		\hline
		\textbf{Uthmani Script} & \textbf{Phonetic Script} \\
		\hline
		\arb{وَقَالَا ٱلْحَمْدُ}       & \arb{وَقَاالَ لحَمدڇ}        \\
		\hline
		\arb{نُوحٌ ٱبْنَهُۥ}         & \arb{نُۥۥحُنِ بڇنَه}         \\
		\hline
	\end{tabular}
\end{table*}

\subsubsection{Shadda (\arb{التشديد})}

Shadda (\arb{ّ}) indicates that a letter is doubled or geminated. We represent this by repeating the letter twice, as shown in \ref{tab:ex_tasheel}.

\subsubsection{Pausing (\arb{الوقف})}

Several rules apply at pause (\arb{وقف}):

\begin{itemize}
	\item Vowels (harakat) such as fatha, damma, and kasra are elided, meaning the final letter becomes sakin (\arb{ساكن}).
	\item Small Silah Madd is elided.
	\item Taa marboota (\arb{ة}) is converted to haa (\arb{ه}).
\end{itemize}

\subsubsection{Qalqala (\arb{القلقة})}

Qalqala (\arb{قلقة}) is defined in tajweed as: "a small sound is followed by on one the letter (\arb{ق - ط - ب - ج - د}) if one of them is sakin (\arb{ساكن}) either in between words (\arb{وصلا}) or at pause (\arb{وقفا})" \cite{AlHamad2008}. We denote this small sound as (\arb{ڇ}) like in table \ref{tab:ex_two_saken}.

\subsubsection{Imala (\arb{الإمالة})}

Imala (\arb{إمالة}) is defined in Tajweed as "pronouncing a fatha somewhere between a fatha and a kasra, and an alif somewhere between an alif and a yaa" \cite{sweed2021}. We denote a fatha with imala as \texttt{fatha\_momala} (\arb{۪}) and an alif with imala with two \texttt{alif\_momala} symbols (\arb{ــ}), similar to the representation of Normal Madd. Table \ref{tab:ex_imala} provides an example:

\begin{table*}[ht]
	\centering
	\caption{The table shows how we represent fatha with imala as (\arb{۪}) and alif with imala as (\arb{ــ}). The letter jeem (\arb{ج}) also exhibits qalqala, denoted by (\arb{ڇ}).}
	\label{tab:ex_imala}
	\begin{tabular}{|c|c|}
		\hline
		\textbf{Uthmani Script} & \textbf{Phonetic Script} \\
		\hline
		\arb{مَجْر۪ىٰهَا}            & \arb{مَجڇر۪ــهَاا}          \\
		\hline
	\end{tabular}
\end{table*}

\subsubsection{Tasheel (\arb{التسهيل})}

Tasheel is defined in Tajweed as "pronouncing a hamza (\arb{ء}) with a quality intermediate between a full hamza and the following madd letter, similar to an intermediate vowel (\arb{حركة}) between fatha, damma, and kasra" \cite{sweed2021}. We denote this facilitated hamza with the symbol \texttt{hamza\_mosahala} (\arb{ٲ}). Table \ref{tab:ex_tasheel} provides an example:

\begin{table*}[ht]
	\centering
	\caption{The table shows a hamza with Tasheel denoted by (\arb{ٲ}), along with the disassembly of the letter yaa (\arb{ي}) with shaddah (\arb{ّ}) into two yaa symbols.}
	\label{tab:ex_tasheel}
	\begin{tabular}{|c|c|}
		\hline
		\textbf{Uthmani Script} & \textbf{Phonetic Script} \\
		\hline
		\arb{ءَا۬عْجَمِىٌّ}            & \arb{ءَٲعجَمِيي}            \\
		\hline
	\end{tabular}
\end{table*}

\subsubsection{Sakt (\arb{السكت})}

Sakt is defined in tajweed by "cutting voice without releasing of breathe for short period learned from expert reciters" \cite{AlHamad2008}. Sakat happens in a specified positions in the Hafs recitation. We denote sakt by \texttt{sakt} (\arb{ۜ}).

\end{document}